\documentclass[twocolumn,showpacs,preprintnumbers,amsmath,amssymb,superscriptaddress]{revtex4}
\usepackage{graphicx}% Include figure files
\usepackage{dcolumn}% Align table columns on decimal point
\usepackage{bm}% bold math
\usepackage[chatter]{rotating}

\def\bea{\begin{eqnarray}}
\def\eea{\end{eqnarray}}

\newenvironment{color}[3]{
}{
}

\newcommand{\cyan}[1]      {\begin{color}{0}{1}{1}{#1}\end{colJor}}

\begin{document}

%\preprint{Version 1.5}

\title{Phenomenological analysis of angular correlations in 7 TeV
proton-proton collisions from the CMS experiment}

\author{R. L. Ray} 
\affiliation{Department of Physics, The University of Texas, Austin, Texas 78712 USA}

%%%%%%%%%%%%%%%%%%%%%%%%%%%%%%%%%%%%%%%
\date{\today}

\begin{abstract}
A phenomenological analysis is presented of recent two-particle angular correlation data on
relative pseudorapidity ($\eta$) and azimuth reported by the Compact Muon
Solenoid (CMS) Collaboration for $\sqrt{s}$ = 7~TeV proton-proton collisions.
The data are described with an empirical jet-like
model developed for similar angular correlation measurements obtained from heavy ion collisions at
the Relativistic Heavy Ion Collider
(RHIC).  The same-side (small relative azimuth), $\eta$-extended correlation
structure, referred to as the {\em ridge}, is compared with
three phenomenological correlation structures suggested by theoretical 
analysis.  These include additional angular correlations due to soft gluon radiation in $2 \rightarrow 3$ partonic processes,
a one-dimensional same-side correlation ridge on azimuth motivated for example by color-glass condensate models,
and an azimuth quadrupole similar to that required to describe heavy ion angular correlations.
The quadrupole model provides the best overall description of the CMS data, including the ridge, based on $\chi^2$
minimization in agreement with previous studies. Implications of these results
with respect to possible mechanisms for producing the CMS same-side
correlation ridge are discussed.
\end{abstract}


\maketitle

%%%%%%%%%%%%%%%%%%%%%%%%%%%%%%%%%%%%%%%%%%%%%%%%%%%%%%%%%%%%%%%%%%%%%%%%%%%
\section{Introduction}
\label{Sec:I}

%%%%%%%%%%%%%%%%%%%%%%%%%%%%%%%%%%%%%%%%%%%%%%%%%%%%%%%%%%%%%%%%%%%%%%%%%%%

The CMS Collaboration at the Large Hadron Collider (LHC) at CERN recently
reported the observation of long-range two-particle angular correlations on relative
pseudorapidity ($\eta_\Delta \equiv \eta_1 - \eta_2$) for charged particle pairs at mid-rapidity
with relative azimuth ($\phi_\Delta \equiv \phi_1 - \phi_2$) $< \pi/2$ (near-side or same-side - SS)
in high multiplicity proton-proton
(p-p) collisions at $\sqrt{s}$ = 7 TeV~\cite{CMS}. The collaboration
noted the visual similarity of this same-side ``ridge'' with two-particle
angular correlation structures observed in heavy ion collisions at
RHIC~\cite{Ayaminijet,MikeQM08,JoernRidge,PHOBOSRidge}.
For the latter, two-dimensional (2D) angular correlation analysis reveals at
least two structures which produce same-side, relative $\eta$ elongated correlations:
(1) a quadrupole, $\cos{2\phi_\Delta}$, independent of $\eta$ (within two
units at mid-rapidity) whose amplitude is approximately proportional to the 
square of the eccentricity of the transverse distribution of colliding nucleons from the
impacting nuclei~\cite{DavidHQ}, and (2) a 2D peak centered at ($\eta_\Delta,\phi_\Delta$)
= (0,0) whose amplitude and width on relative pseudorapidity increase markedly
with more central collisions (i.e. reduced impact parameter)~\cite{MikeQM08}.
The quadrupole correlation in heavy ion collisions is conventionally attributed to pressure driven,
hydrodynamic flow~\cite{hydro} due to rapid equilibration of a partonic medium~\cite{Mueller}.
Inconsistencies between this assumed scenario and other RHIC data have been
found however~\cite{Tomv2bash}; alternate mechanisms for quadrupole correlation production have been studied~\cite{Boreskov,Kopeliovich}.
The origin of the $\eta$-elongated 2D peak at RHIC is not understood although many ideas have been
proposed.
The appearance of one or both of these ridge-like correlation structures in p-p collisions
at the LHC would be quite interesting.

A number of mechanisms have been proposed to explain the $\eta$-elongated 2D
correlation peak observed in the heavy ion collision data at RHIC. Several of these authors
have extended their models to include the CMS p-p results. Dumitru {\em et al.}~\cite{Dumitru}
proposed that flux tubes of color glass condensate~\cite{CGC} formed immediately after impact
produce correlations among the hadron fragments which are transported
outward by an assumed radial flow. Shuryak~\cite{Shuryak} and Voloshin~\cite{Voloshin} invoked similar ideas
except the initial stage correlations result from beam-jets, which are the fragmentation results of
forward going nucleon remnants following semi-hard partonic scattering. Hwa {\em et al.}~\cite{Hwa}
proposed that the same-side ridge is generated by fast parton (from semi-hard
scattering) - soft, ``thermal'' parton recombination into final-state hadrons.
Wong~\cite{Wong} explained the same-side ridge by way of scattering (momentum kick)
between fast, transverse scattered partons and soft partons of a medium.
Levin and Rezaeian~\cite{Levin} argued that the long-range rapidity and azimuth
correlations between charged hadron pairs produced from two BFKL parton showers will
contribute to the CMS ridge.
Werner {\em et al.}~\cite{Werner} interpreted the CMS results in terms of energy
density fluctuations in the initial collision stage which manifest in the final-state
via hydrodynamic expansion; the relevant correlation being a quadrupole.

Within perturbative QCD Field~\cite{Field} suggested that
initial and final state radiation effects in partonic
$2 \rightarrow 3$ and $2 \rightarrow 4$ processes
may contribute to extending the $\eta$ range of the
jet correlations.  Sj\"{o}strand~\cite{Sjostrand} and Trainor~\cite{Tomfrag}
independently suggested that fragmenting color flux tubes, as in the LUND model of soft particle
production~\cite{LUND}, stretching between transverse scattered colored partons
and longitudinal colored nucleon remnants from semi-hard collisions could produce
$\eta$ elongation in the jet correlation peak. None of the preceding models have
been shown to account for all of the correlation~\cite{RayCERP}  and $p_t$ spectrum~\cite{Tomspec} features in the
heavy-ion data from RHIC which are associated with the $\eta$-elongated peak
phenomenon.

In a physical model-independent analysis Trainor and Kettler~\cite{Tomridge} showed that
the systematic dependences of the quadrupole correlation amplitude on collision
centrality, event multiplicity, and collision energy ($\sim \log{\sqrt{s}}$)
determined from RHIC heavy-ion data~\cite{DavidHQ}, when extrapolated to LHC energies for high multiplicity p-p
collision events, predicted quadrupole amplitudes which
were in reasonable agreement with the CMS near-side ridge.
They suggested that the CMS ridge and the quadrupole correlations at RHIC
have a common dynamical origin, but not necessarily hydrodynamic. Bo\.zek~\cite{Bozek}
also showed that the transverse momentum ($p_t$) selected azimuth projections of
the CMS 2D correlations at larger relative pseudorapidity~\cite{CMS} were well described
with $\cos{\phi_\Delta}$ (dipole) and $\cos{2\phi_\Delta}$ (quadrupole) terms only.

In the present analysis the 7 TeV p-p angular correlation data
from the CMS experiment were fitted in 2D angular space ($\eta_\Delta,\phi_\Delta$)
using a multi-component model similar to that used to describe angular correlations
from Au-Au collisions at RHIC~\cite{MikeQM08}. Four data sets were studied
corresponding to combinations of event selection (minimum-bias or high
multiplicity) and particle $p_t$ selection ($p_t > 0.1$~GeV/$c$ or 
$1 < p_t < 3$~GeV/$c$). Additional components (soft gluon radiation, azimuth quadrupole and same-side Gaussian ridge on azimuth) were included in the
fitting model in order to phenomenologically
represent the correlation structures expected for each class of theoretical models
listed above. 
The purpose of this analysis is to determine which
phenomenological description(s) of the p-p CMS correlation data is preferred
based on fit quality
in an effort to guide ongoing theoretical studies of the CMS ridge.

The paper is organized as follows.
The phenomenological fitting model is explained in the next section;
fitting results are presented in Sec.~\ref{Sec:III}.
The relation of this analysis to previous studies of these data and the implications
of the fitting phenomenology for theoretical models of $\eta$-elongated
angular correlations in high energy collisions are
discussed in Sec.~\ref{Sec:IV}.
A summary and conclusions are given in Sec.~\ref{Sec:V}.

%%%%%%%%%%%%%%%%%%%%%%%%%%%%%%%%%%%%%%%%%%%%%%%%%%%%%%%%%%%%%%%%%%%%%%%%%%%
\section{Angular correlations and fitting model}
\label{Sec:II}
%%%%%%%%%%%%%%%%%%%%%%%%%%%%%%%%%%%%%%%%%%%%%%%%%%%%%%%%%%%%%%%%%%%%%%%%%%%

\subsection{Minimum-bias angular correlations}

The two-particle angular correlation distributions from CMS~\cite{CMS} and
the STAR experiment at RHIC~\cite{Ayaminijet,MikeQM08,JoernRidge} represent
inclusive sums of all charged particle pairs within the tracking acceptances of the 
respective detectors.
There is no special leading particle or ``trigger'' particle in this type of
analysis and in this sense the CMS angular correlations may be referred to as {\em minimum-bias}.
Correlations are obtained from normalized,
binned ratios of real to mixed-event pairs, the latter being constructed by
taking the two particles in a pair from different but similar events.
If events produce more than one jet within the acceptance all intra-jet
charged particle pairs contribute to the correlation peak at small relative angles near (0,0).
Inter-jet two-particle correlations are indistinguishable from background
for uncorrelated jets. Particle pairs from separate jets of a correlated,
back-to-back dijet appear as a broad, opposite-side or away-side ($\phi_\Delta > \pi/2$, AS)
ridge on azimuth centered at $\pi$. 
The same-side $\eta$-elongated correlations may or may not be associated with
jets; proximity of correlation structures on relative angle does not necessarily imply proximity in absolute ($\eta,\phi$) space.

\subsection{Nominal fitting model components}

The primary elements of the present fitting function were developed previously by the
STAR Collaboration in order to describe 2D angular
correlation data for charged particle production from minimum-bias p-p
collisions at $\sqrt{s}$ = 0.2~TeV~\cite{Jeffpp}. Those studies used $p_t$ cuts and
like-sign and unlike-sign pair combinations to reveal simple geometrical
shapes in the correlations.
Those structures include: (1) a jet-like, 2D Gaussian peak at
small relative angles with an accompanying away-side,
$\eta_\Delta$-independent ridge corresponding to approximately back-to-back dijets (e.g. {\sc pythia} predictions in \cite{CMS}) or
other $p_t$ conserving correlation effects, 
(2) a narrower 2D exponential peak at zero opening angle representing quantum
correlations~\cite{HBT} among identical charged particles, and (3) a 1D
Gaussian ridge on $\eta_\Delta$ which is the 2D manifestation
of charge-ordering first observed on relative $\eta$ in collisions at the
ISR~\cite{ISR}. The latter is thought to represent longitudinal (beam
direction) fragmentation as described for example in the LUND model~\cite{LUND}. With the addition of an $\eta$-independent
quadrupole this phenomenological model
quantitatively describes the 2D angular correlation data from minimum-bias
Au-Au collisions at STAR~\cite{Ayaminijet,MikeQM08}.
 
\subsection{Color coherence in $2 \rightarrow 3$ processes}

In addition to the above model components important terms could arise from
color coherence effects~\cite{Ellis,Jaques}.
Soft gluon radiation amplitudes from color connected partons in hard scattering
$2 \rightarrow 2$ processes interfere producing increased yields in the hard
scattering plane defined by the incident and scattered partons. 
Assuming hadronization does not obliterate the soft gluon distribution
the resulting final state hadrons will increase the particle yield near the
accompanying jet with a larger increase occurring along the beam direction, or on
relative pseudorapidity than on relative azimuth.
This next-to-leading-order process may therefore contribute
to the CMS ridge.
For dijets within the
acceptance soft gluon radiation will also contribute to the away-side correlation
ridge on azimuth.

Color coherence effects can be approximated within the soft
gluon momemtum limit ($k \rightarrow 0$)~\cite{Ellis} where the squared
amplitude for the $2 \rightarrow 3$ process is factored into the
product of the probability
density for the $2 \rightarrow 2$ hard scattering times the soft gluon radiation
probability distributions for each color connected parton pair. The total yield is 
obtained by summing over all color connected parton pairs. The dominant $2 \rightarrow 3$
reaction considered here for midrapidity hadron production is
gluon-gluon scattering given by~\cite{Ellis}
\bea
g(p_1) + g(p_2) & \rightarrow & g(p_3) + g(p_4) + g(k),
\label{Eq0}
\eea
where the soft gluon yield is approximately
\bea
P_g & \approx & C_A \sum_{i<j=1}^4 W_{ij},
\label{Eq01}
\eea
the summation includes all parton pairs, gluon color factor $C_A = N_c$
(number of colors), and
\bea
W_{ij} & = & g^2 \frac{p_i \cdot p_j}{p_i \cdot k\, k \cdot p_j}.
\label{Eq02}
\eea
In Eq.~(\ref{Eq02}) $g$ is the strong coupling constant and $p_i \cdot p_j$
etc. denote scalar products of 4-momentum vectors. Quantities $W_{ij}$
are singular for massless partons in the limits $k \rightarrow 0$, $\theta_{ik} \rightarrow 0$,
or $\theta_{jk} \rightarrow 0$ where $\theta$ are the angles between the
parton and radiated soft gluon. To regulate the singularities small cut-off masses (5 MeV/$c^2$) were
assumed for each parton and a Gaussian angular cut-off,
$F_i = [1-{\rm exp}(-\theta_{ik}^2 /2\sigma_g^2)]$ where $\sigma_g = 0.2$~radians,
was applied to each amplitude and $W_{ij}$ was weighted with $F_i F_j W_{ij}$.
The latter cut-off also respects angular ordering requirements for the
subsequent jet fragmentation~\cite{Ellis}.
The soft gluon radiation yield in the soft limit was
approximated by
\bea
P_g & \approx & A_g \frac{g^2 C_A}{E^2_{T_k}} \sin^2\theta_{1k}
\sum_{i<j=1}^4 F_i F_j \frac{\alpha_{ij}}{\alpha_{ik} \, \alpha_{jk}}~,
\label{Eq03}
\eea
where $\alpha_{ij} = 1 - \beta_i \beta_j \cos\theta_{ij}$, $\beta = p/E$
(velocity), gluon 1 defines the beam direction, $E_{T_k}$ is the 
transverse energy of the radiated gluon, and $A_g$ is a phenomenological amplitude
which will be adjusted by the fitting program.
The final-state hadron distribution resulting from the soft radiated gluons is 
assumed to be given by Eq.~(\ref{Eq03}) according to the local parton-hadron
duality hypothesis (LPHD)~\cite{LPHD}.

The soft gluon radiation adds to the final-state hadron distribution associated with
minimum-bias jet production, given by $P_{\rm jet,3} + P_{\rm jet,4} + P_g$.
The first two yields correspond to hadron distributions from
the fragmentation of hard scattered partons 3 and 4 in the reaction in Eq.~(\ref{Eq0}).  Each is represented 
by  a 2D Gaussian distribution
\bea
P_{\rm jet}(\eta,\phi) & = & A e^{-(\eta - \eta_{\rm jet})^2 / 2 \sigma^2_\eta} e^{-(\phi - \phi_{\rm jet})^2 / 2 \sigma^2_\phi},
\label{Eq04}
\eea
where $\eta,\phi$ are defined within the detector acceptance.
The single particle Gaussian widths are directly proportional to the widths of
the jet-like angular correlation
on $\eta_\Delta$ and $\phi_\Delta$ discussed below.
It is expected
that $P_g \ll P_{\rm jet}$ and the additional contribution to the 2D
angular correlations is dominated by the cross term
$P_g(P_{\rm jet,3} + P_{\rm jet,4})$ projected onto relative azimuth and
pseudorapidity and averaged over all jet angular positions within the
acceptance. The 2D angular autocorrelation in the soft gluon
radiation limit for $2 \rightarrow 3$ processes was approximated by
\begin{widetext}
\bea
A_{\rm jet,soft-g}(\eta_\Delta,\phi_\Delta) & \approx & \frac{1}{8\Omega(\eta_\Delta)}
\frac{1}{2\pi \Delta\eta} \int_{-\pi}^\pi d\phi_{\rm jet,3}
\int_{-\Delta\eta/2}^{\Delta\eta/2} d\eta_{\rm jet,3}
\int_{-\Omega(\eta_\Delta)}^{\Omega(\eta_\Delta)} d\eta_\Sigma
\int_{-\pi}^\pi d\phi_\Sigma \nonumber \\
 & \times & \left\{ P_g(\eta_1,\phi_1) 
 \left[ P_{\rm jet,3}(\eta_2,\phi_2)+P_{\rm jet,4}(\eta_2,\phi_2) \right] 
 +   P_g(\eta_2,\phi_2)
\left[ P_{\rm jet,3}(\eta_1,\phi_1)+P_{\rm jet,4}(\eta_1,\phi_1)\right] \right\}
\label{Eq05}
\eea
\end{widetext}
where $\Delta\eta$ is the (symmetric) pseudorapidity acceptance,
$\Omega(\eta_\Delta) = |\Delta\eta - |\eta_\Delta ||$,
$\eta_\Sigma = \eta_1 + \eta_2$ and $\phi_\Sigma = \phi_1 + \phi_2$. Jets
3 and 4 were for simplicity assumed to be uniformly distributed within the
$2\pi\Delta\eta$ acceptance and back-to-back in the p-p collision c.m.
system (CMS lab frame), where $\phi_{\rm jet,4} = \phi_{\rm jet,3} \pm \pi$
and $\eta_{\rm jet,4} = - \eta_{\rm jet,3}$. $P_g(\eta,\phi)$ was obtained
from Eq.~(\ref{Eq03}) for back-to-back dijets at angular positions
$(\eta_{\rm jet,3},\phi_{\rm jet,3})$ and $(\eta_{\rm jet,4},\phi_{\rm jet,4})$
with the soft gluon (resulting final-state hadron) at $(\eta,\phi)$.

\begin{figure}[t]
\includegraphics[keepaspectratio,width=1.65in]{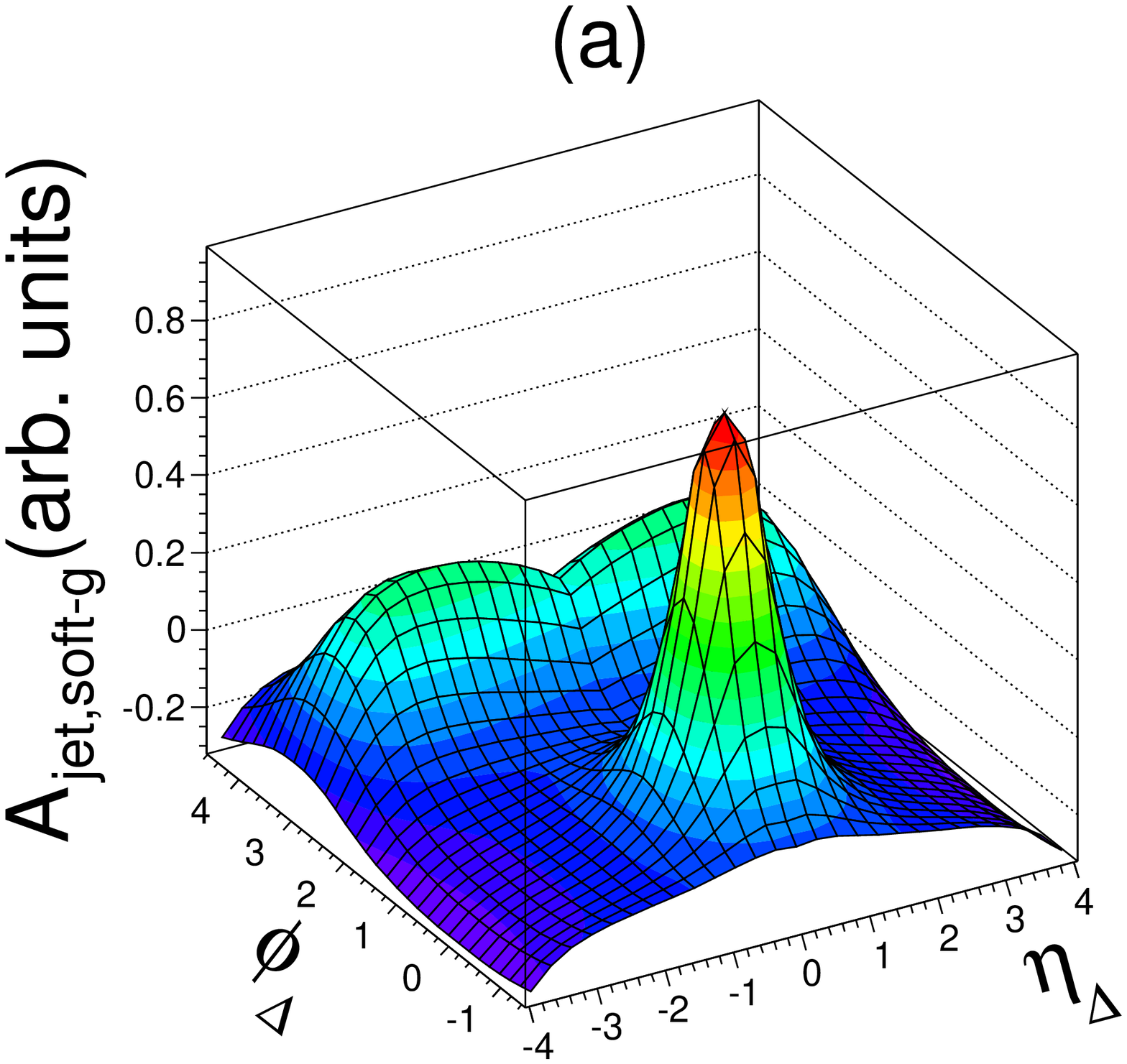}
\includegraphics[keepaspectratio,width=1.65in]{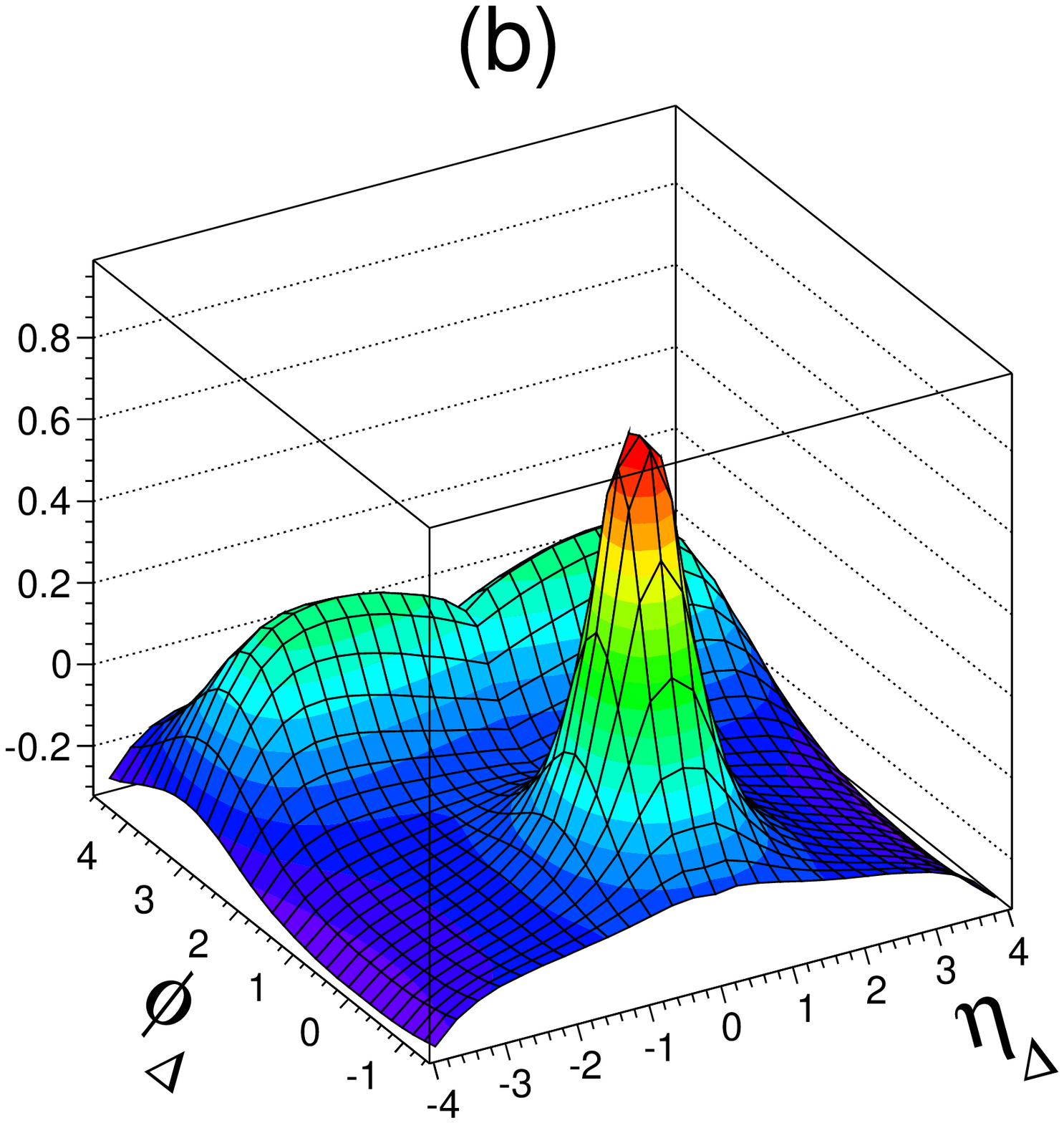}
\includegraphics[keepaspectratio,width=1.65in]{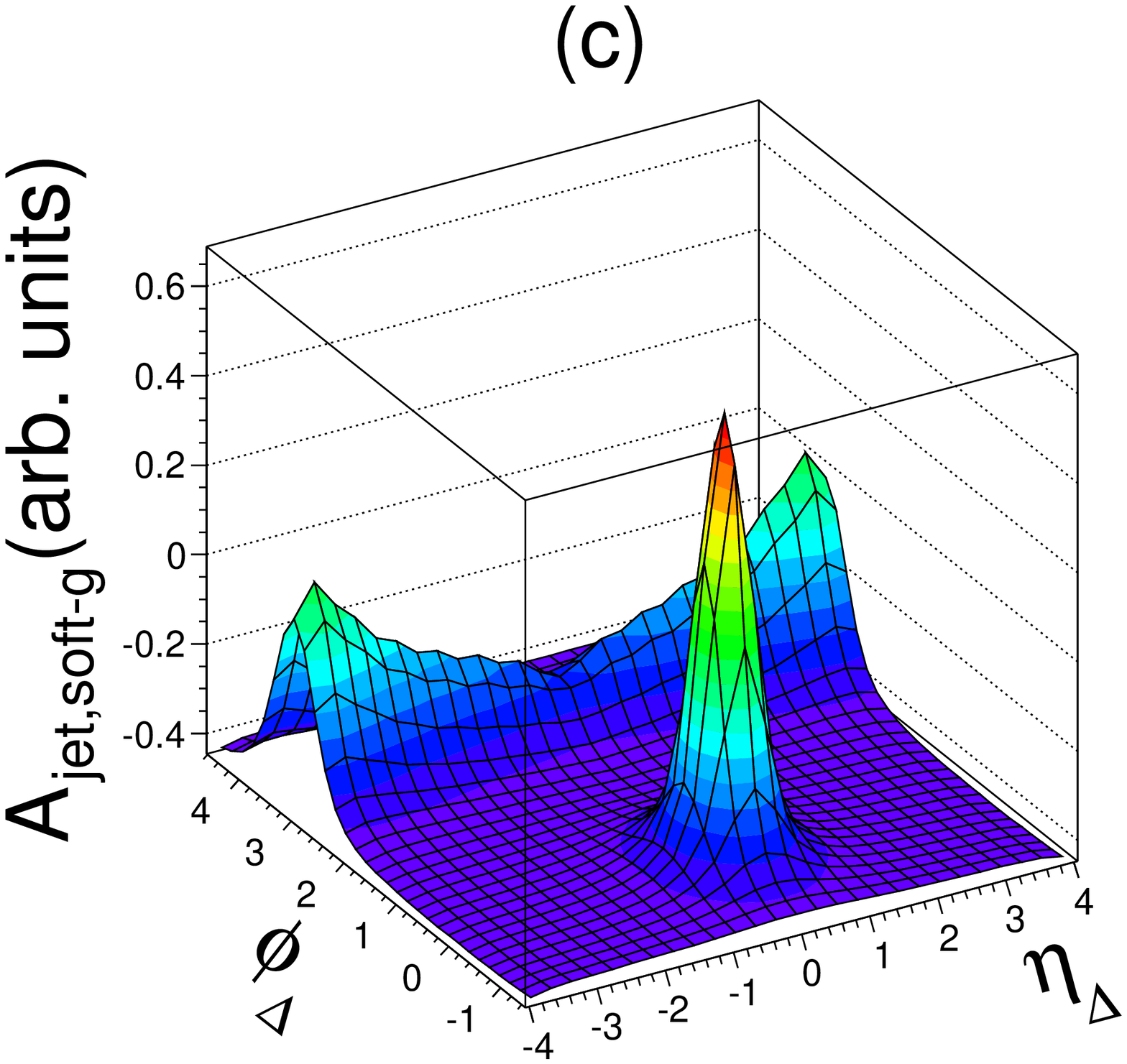}
\includegraphics[keepaspectratio,width=1.65in]{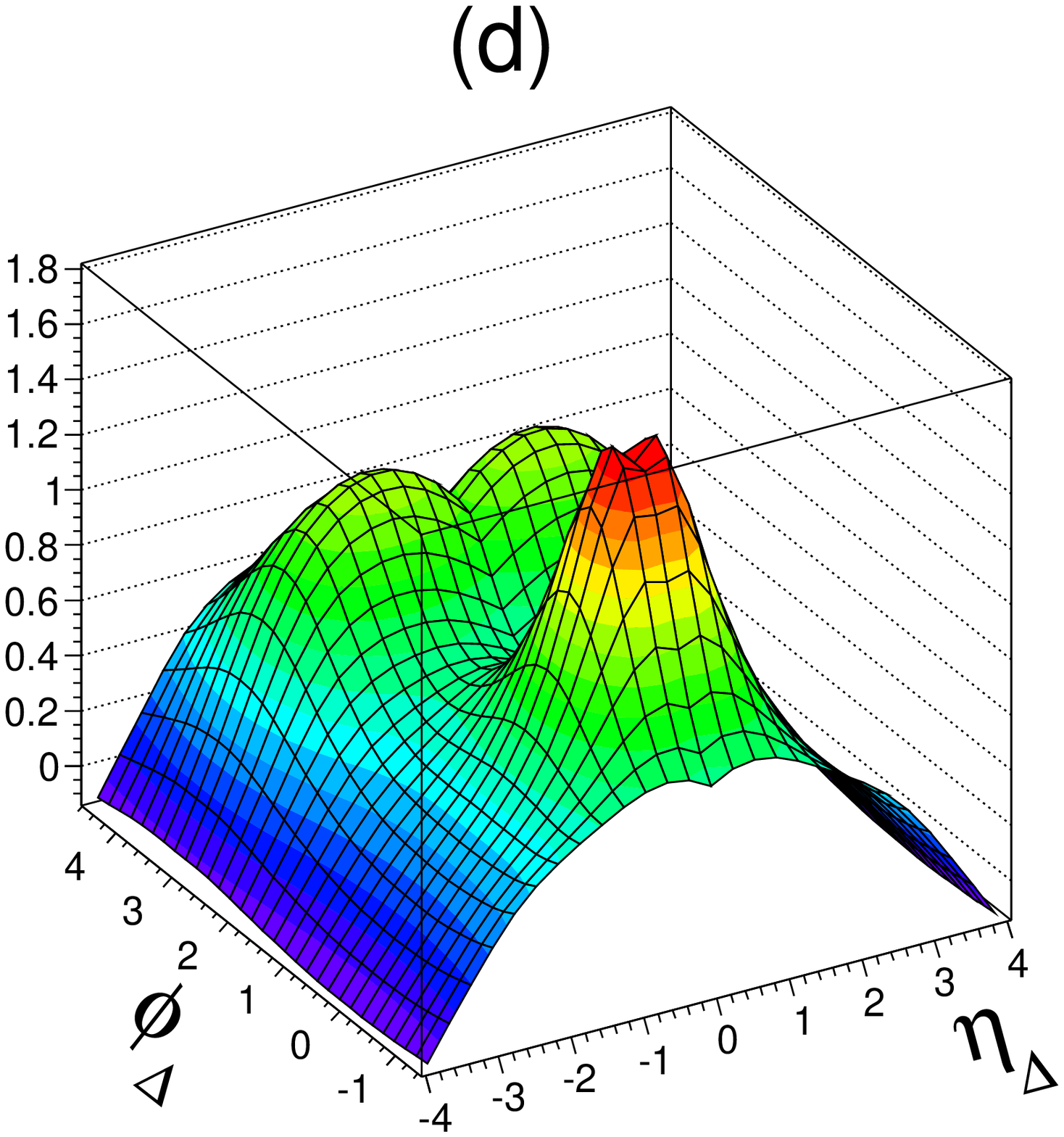}
\caption{\label{Fig0}  
(Color online) Perspective views of two-dimensional charge-independent angular
correlations from Eq.~(\ref{Eq05}), assuming arbitrary normalization, for soft-gluon radiation from back-to-back
Gaussian jets estimated
in the soft limit ($k \rightarrow 0$) on relative azimuth and pseudorapidity
$(\eta_\Delta,\phi_\Delta)$ for the high multiplicity, $1 < p_t < 3$~GeV/$c$ data from CMS. 
From upper-left, to lower-right panels (a)-(d) show angular correlations
assuming
nominal cut-off masses 5~MeV and angle 0.2 rad; 1~MeV, 0.2 rad; 5~MeV, 0.05 rad;
 and 5~MeV, 0.35 rad, respectively. The single particle jet distribution widths
on $\eta$ and $\phi$ were 0.20 and 0.22 respectively.}
\end{figure}

The resulting angular correlations from Eq.~(\ref{Eq05}) for the high multiplicity,
$1 < p_t < 3$~GeV/$c$ data set are shown in Fig.~\ref{Fig0}
for different values of cut-off parameters and for jet widths
$\sigma_\eta$ = 0.20 and $\sigma_\phi$ = 0.22. The parton and soft gluon
cut-off masses and collinear cut-off $\sigma_g$ assumed in panels (a) - (d)
were 5~MeV and 0.2 radians, 1~MeV and 0.2 radians, 5~MeV and 0.05 radians,
and 5~MeV and 0.35 radians, respectively. In each panel the amplitude was
arbitrarily adjusted to unit height difference between the positions
on $(\eta_\Delta,\phi_\Delta)$ at $(0,0)$ and $(0,\pi/2)$.
The away-side
ridge corresponds to correlations between soft-gluons emitted near jet-3
and the final-state hadrons from jet-4 and vice versa. The increased amplitude
of the away-side ridge at larger $|\eta_\Delta|$ results from the increased
radiative yields when the angles between incident and outgoing partons
decrease. The fall-off at the largest $|\eta_\Delta|$ results from both the
collinear cut-off $\sigma_g$ and the $\sin^2\theta_{1k}$ factor in Eq.(\ref{Eq03}).

The angular correlations are weakly sensitive to the assumed cut-off mass
(panels a and b), but
for masses of order the pion mass the correlations become similar to that in
panel (d). The angular correlations are however
strongly affected by the collinear cut-off. For the fits to data
$\sigma_g = 0.2$ was selected (panel a) in consideration of angular
ordering while minimizing the
suppression at larger $|\eta_\Delta|$. For the fits to the four 7~TeV data sets
cut-off parameters 5~MeV and 0.2 radians were always used. The jet single
particle Gaussian widths on $\eta$ and $\phi$ were (0.35,0.52), (0.22,0.26),
(0.30,0.30) and (0.20,0.22) for the minimum-bias low and higher $p_t$ data,
and the high multiplicity, low $p_t$ and higher $p_t$ data sets, respectively.

%%%%%%%%%%%%%%%%%%%%%%%%%%%%%%%%%%
%%%%%%%%%%%%%%%%%%%%%%%%%%%%%%%%%%

\subsection{Ridge components}

Recent theoretical predictions~\cite{Werner,Casalderrey}  
and phenomenological analysis of the 7 TeV data from CMS
\cite{Tomridge,Bozek} warrant including the quadrupole component.
Although the presence of the quadrupole correlation in the charged particle
yields from relativistic heavy ion collisions is often cited as evidence
for hydrodynamic evolution, other theoretical studies~\cite{Boreskov,Kopeliovich}
show that a quadrupole correlation may result directly from QCD.
%Jet correlation shape modification due to higher twist contributions~\cite{Field} or additional Lund-model color flux
%tube fragmentation~\cite{Sjostrand,Tomfrag} would produce non-Gaussian, perhaps exponential shapes
%in addition to independent $\eta$ and $\phi$ width changes.

Models assuming outward, collective flow with initial-state energy fluctuations ({\em e.g.} CGC, beam-jets)
or recombination suggest including a same-side
1D ridge on azimuth with no accompanying away-side structure other than that
already present in the model.  
Momentum conservation induced correlations associated with this class of
models would manifest as an away-side ridge on $\eta_\Delta$ and are therefore
already included in the fitting function.

\subsection{Specific modifications for the CMS data}

Other modifications of the fitting model in~\cite{MikeQM08} were required 
for the CMS data.  
At RHIC energies the away-side correlation ridge (non-quadrupole) for lower $p_t$
particles is well described by a dipole, $\cos{\phi_\Delta}$, 
which is understood to represent the limit of a periodic away-side azimuth
ridge (Gaussian) as its width increases~\cite{footperiod}.
Preliminary studies showed that the 7~TeV CMS correlation data were
better described with a periodic Gaussian.
The measured correlations also display significant away-side $\eta_\Delta$-dependence.
Analysis of Au-Au quadrupole correlations with large $\eta$ acceptance
indicates that reduction in amplitude with increasing $|\eta_\Delta|$ is possible~\cite{v2eta}.
Consequently, the away-side 1D Gaussian, the same-side 1D Gaussian,
and the quadrupole components all include symmetric $\eta_\Delta$-dependent
modulations.

\subsection{Fitting model}
\label{Sec:FitModel}

The complete fitting
function used here is given by~\cite{footambig},
\bea
F(\eta_\Delta,\phi_\Delta) & = & A_0 + A_Q f(\eta_\Delta) \cos(2\phi_\Delta)
\nonumber \\
& + & A_1e^{-\frac{1}{2} \left\{ \left( \frac{\eta_\Delta}{\sigma_{\eta_\Delta}} \right)^{2\alpha}
     + \left( \frac{\phi_\Delta}{\sigma_{\phi_\Delta}} \right)^{2\beta} \right\} }
\nonumber \\
 & + & A_2 e^{-\left\{ \left( \frac{\eta_\Delta}{w_\eta} \right)^2 +
                       \left( \frac{\phi_\Delta}{w_\phi} \right)^2 \right\}^{1/2}}
+ A_3 e^{-\frac{1}{2} \left( \frac{\eta_\Delta}{\sigma_0} \right)^{2\gamma}}
\nonumber \\
 & + & A_{4}\left[ g(\eta_\Delta) F_{AS}(\phi_\Delta) - \left( F_{AS}(0) + F_{AS}(\pi) \right) /2 \right]
\nonumber \\
 & + & A_{5}\left[ f(\eta_\Delta) F_{SS}(\phi_\Delta) - \left( F_{SS}(0) + F_{SS}(\pi) \right) /2 \right],
\nonumber \\
F_{AS}(\phi_\Delta) & = & \sum_{k=odd-int}
e^{-\frac{1}{2} \left( \frac{\phi_\Delta - k\pi}{\sigma_{AS}} \right)^2},
\nonumber \\
F_{SS}(\phi_\Delta) & = & \sum_{k=even-int}
e^{-\frac{1}{2} \left( \frac{\phi_\Delta - k\pi}{\sigma_{SS}} \right)^2},
\label{Eq1}
\eea
where $f(\eta_\Delta) = 1 + \delta \eta_\Delta^2$,
$g(\eta_\Delta) = 1 + \epsilon \eta_\Delta^2 + \zeta \cos(2\pi \eta_\Delta / \Delta \eta)$,
and the summations for the periodic away-side and same-side 1D Gaussian
ridges ($A_4$ and $A_5$ terms) were accurately truncated with sums
($-3,-1 \cdots 5$) and ($-4,-2 \cdots 4$), respectively.
The quadrupole ($A_Q$ term) and the same-side periodic 1D Gaussian ($A_5$
term) were applied alternately, not simultaneously in the fitting.
Terms periodic in $\phi_\Delta$ for the $A_1$ and $A_2$ components [i.e. dependent on
$\phi_\Delta - (\pm 2\pi)$] were included but were negligible for the azimuth widths
used here.  The constant offsets for the AS and SS Gaussian ridge components were
included in order that coefficients $A_4$ and $A_5$ only (approximately) affect the
amplitudes of the $\phi_\Delta$-dependent oscillations and not the overall offset.

Function $F(\eta_\Delta,\phi_\Delta)$ was fitted to measured quantity
$R(\eta_\Delta,\phi_\Delta)$ from CMS~\cite{CMS}.
The fitting parameters
include: $A_0$ (normalization offset), $A_Q$ or $A_5$ and
$\sigma_{SS}$ with $\delta$, same-side jet parameters $A_1$, $\sigma_{\eta_\Delta}$, $\sigma_{\phi_\Delta}$,
$\alpha$ and $\beta$, 2D exponential parameters $A_2$, $w_\eta$ and $w_\phi$,
1D Gaussian parameters $A_3$, $\sigma_0$ and $\gamma$ (only required for the
minimum-bias, $p_t > 0.1$~GeV/c data), and away-side 1D
Gaussian ridge parameters $A_4$, $\sigma_{AS}$, $\epsilon$ and $\zeta$.
Due to the similar magnitudes of the width parameters for the $A_1$ and $A_2$ terms the
fitting model may not provide an accurate separation between jet fragment correlations
and quantum correlations in each case.

The soft gluon radiation component was computed for each case using the
Gaussian width parameters for the jets obtained by fitting the data without
the soft radiation component and with the cut-off parameters discussed in
the above subsection. The single particle jet width parameters in Eq.~(\ref{Eq04}) are related to the SS 2D correlation peak Gaussian widths by
$\sigma_\eta = \sigma_{\eta_\Delta}/\sqrt{2}$ and
$\sigma_\phi = \sigma_{\phi_\Delta}/\sqrt{2}$ owing to the definition of the
difference variables $\eta_\Delta$ and $\phi_\Delta$.
The function computed in Eq.~(\ref{Eq05}) was
arbitrarily adjusted in overall magnitude by the fitting program in order
to obtain the best fit.

\subsection{Fitting errors}

Best fits were selected by $\chi^2$ per degree of freedom (dof), residuals, 
the 2D $\chi^2$ distributions, and parameter fit errors.
The $\chi^2$/dof values were fairly
large indicating that residual, systematic effects are statistically significant.
Such effects may include experimental artifacts and/or physical
processes which are not accurately described by the fitting model.
The fit parameter error covariance matrix was estimated by the following matrix
inversion,
\bea
\delta^2 P_{jk} & \approx & \left[
\sum_i \frac{1}{\varepsilon_i^2} \frac{\partial F_i}{\partial P_j}
\frac{\partial F_i}{\partial P_k} \right] ^{-1}
\label{Eq2}
\eea
where the summation includes all unique $(\eta_\Delta,\phi_\Delta)$ bins,
$\varepsilon_i$ is the statistical error, and $\partial F_i/\partial P_j$
are partial derivatives of the model function with respect to fit parameter
$P_j$ at the $\chi^2$ minimum. This estimate assumes that near the $\chi^2$ minimum
the fitting function dependence on parameter $P_j$ can be accurately expanded to the leading order (linear) terms in a Taylor
series. Parameter fit errors, $\sqrt{\delta^2 P_{ii}}$, are listed along with the best fit values in
Table~\ref{TableI}. The CMS collaboration quotes an overall 15\% systematic
uncertainty in the data near the same-side ridge resulting in an additional
15\% systematic uncertainty in the quadrupole and same-side ridge amplitudes
$A_Q$ and $A_5$.

%%%%%%%%%%%%%%%%%%%%%%%%%%%%%%%%%%%%%%%%%%%%%%%%%%%%%%%%%%%%%%%%%%%%%%%%%%%
\section{Results}
\label{Sec:III}
%%%%%%%%%%%%%%%%%%%%%%%%%%%%%%%%%%%%%%%%%%%%%%%%%%%%%%%%%%%%%%%%%%%%%%%%%%%

\begin{figure*}[t]
\includegraphics[keepaspectratio,width=1.70in]{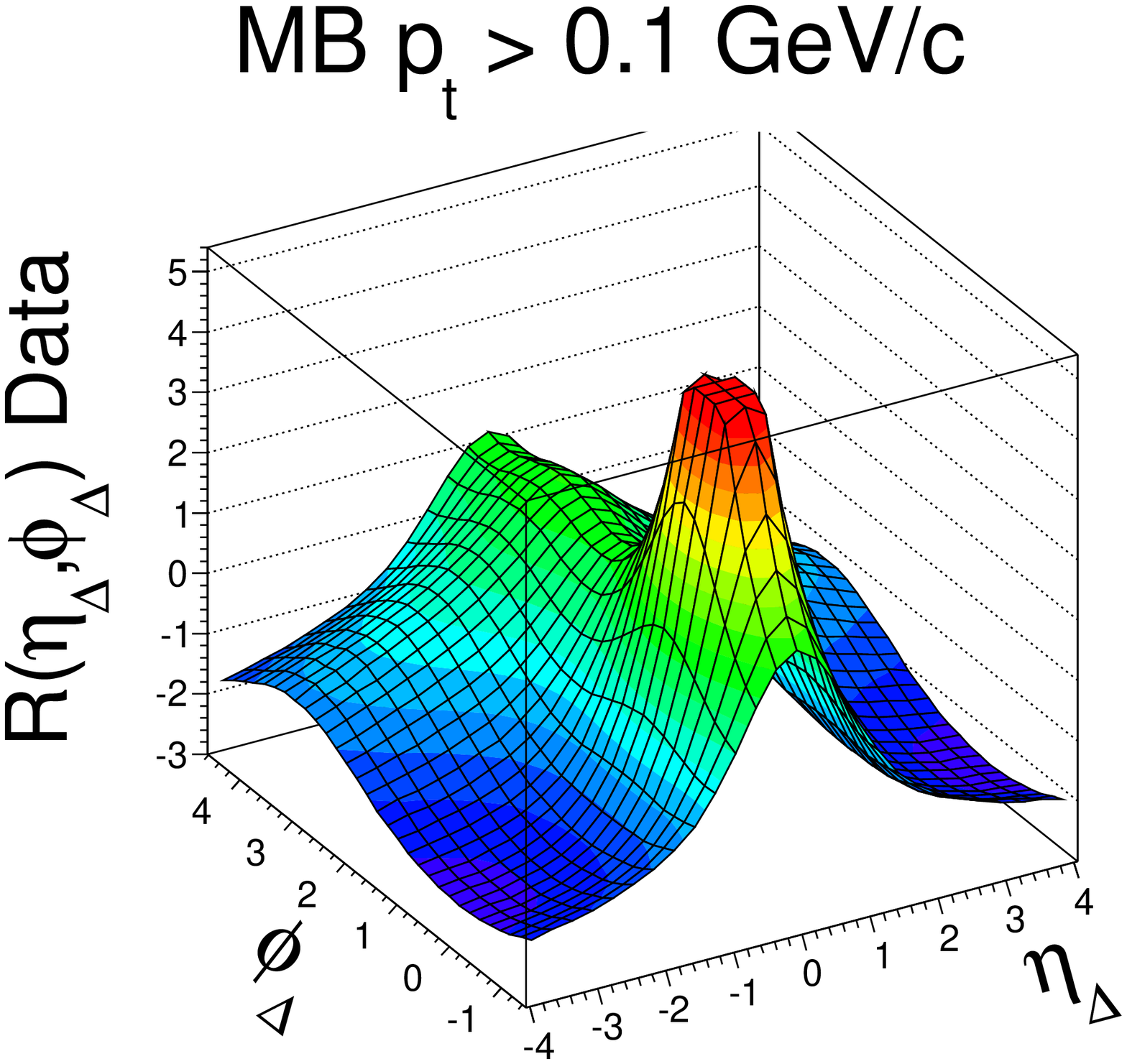}
\includegraphics[keepaspectratio,width=1.70in]{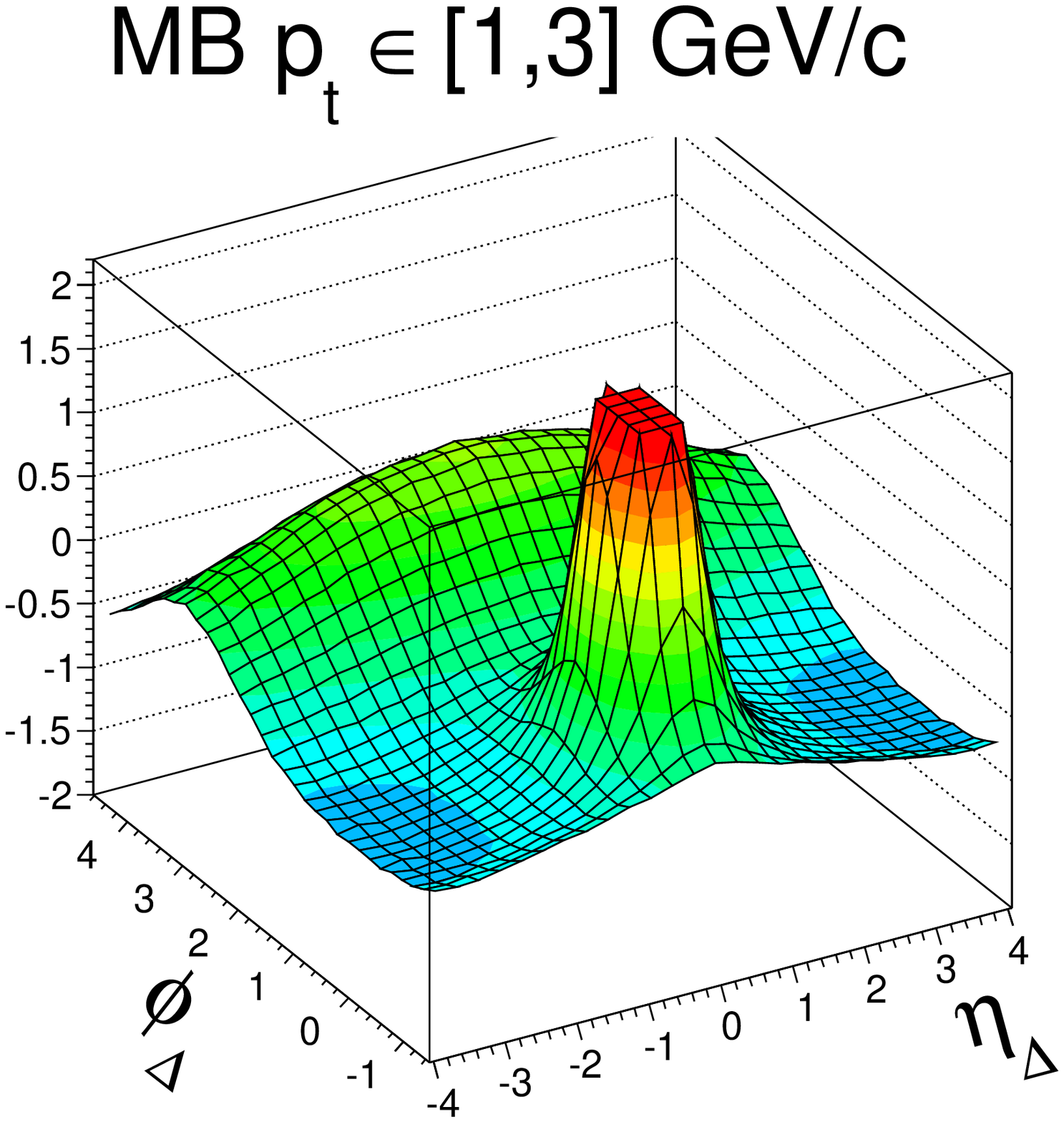}
\includegraphics[keepaspectratio,width=1.70in]{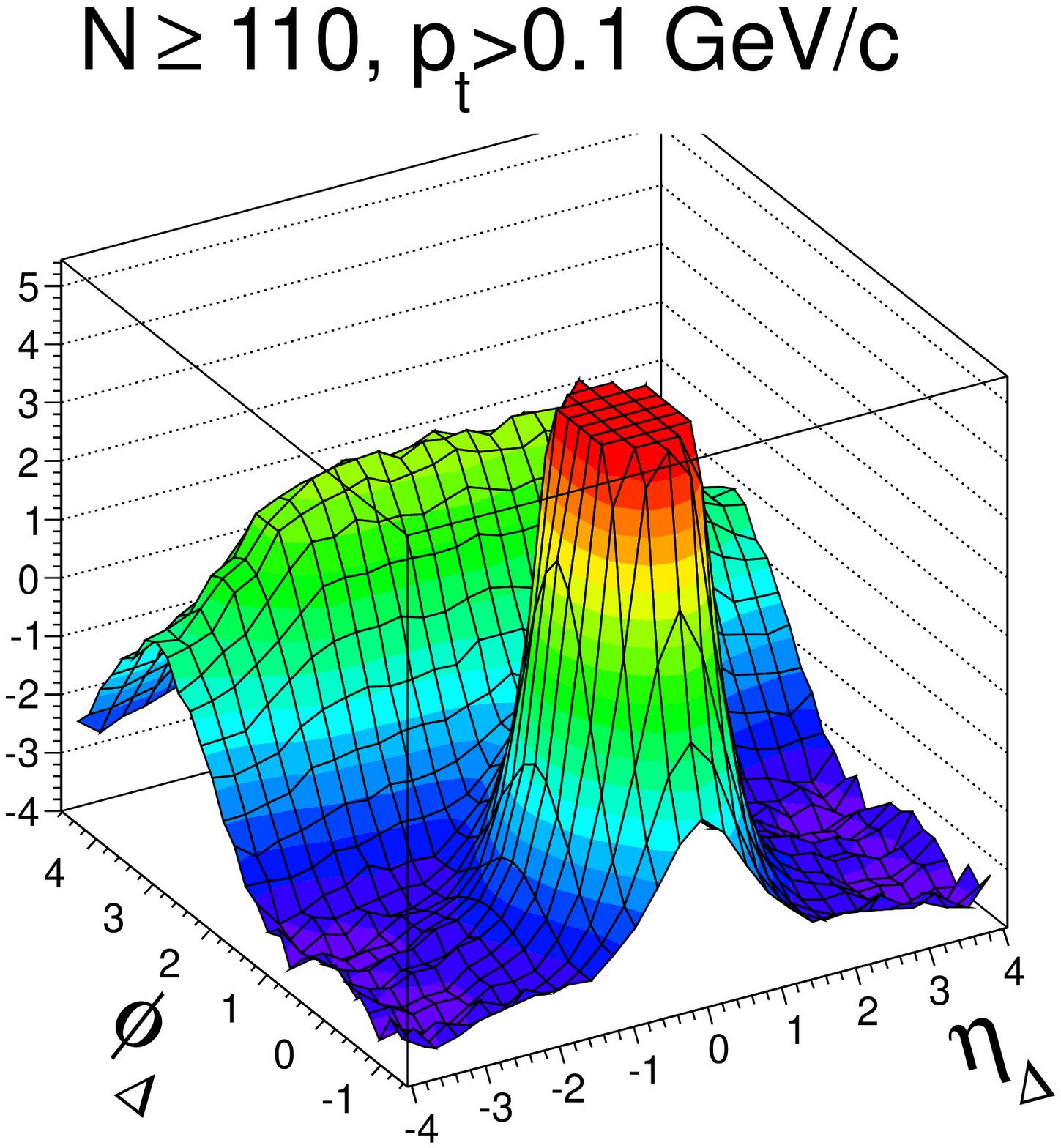}
\includegraphics[keepaspectratio,width=1.70in]{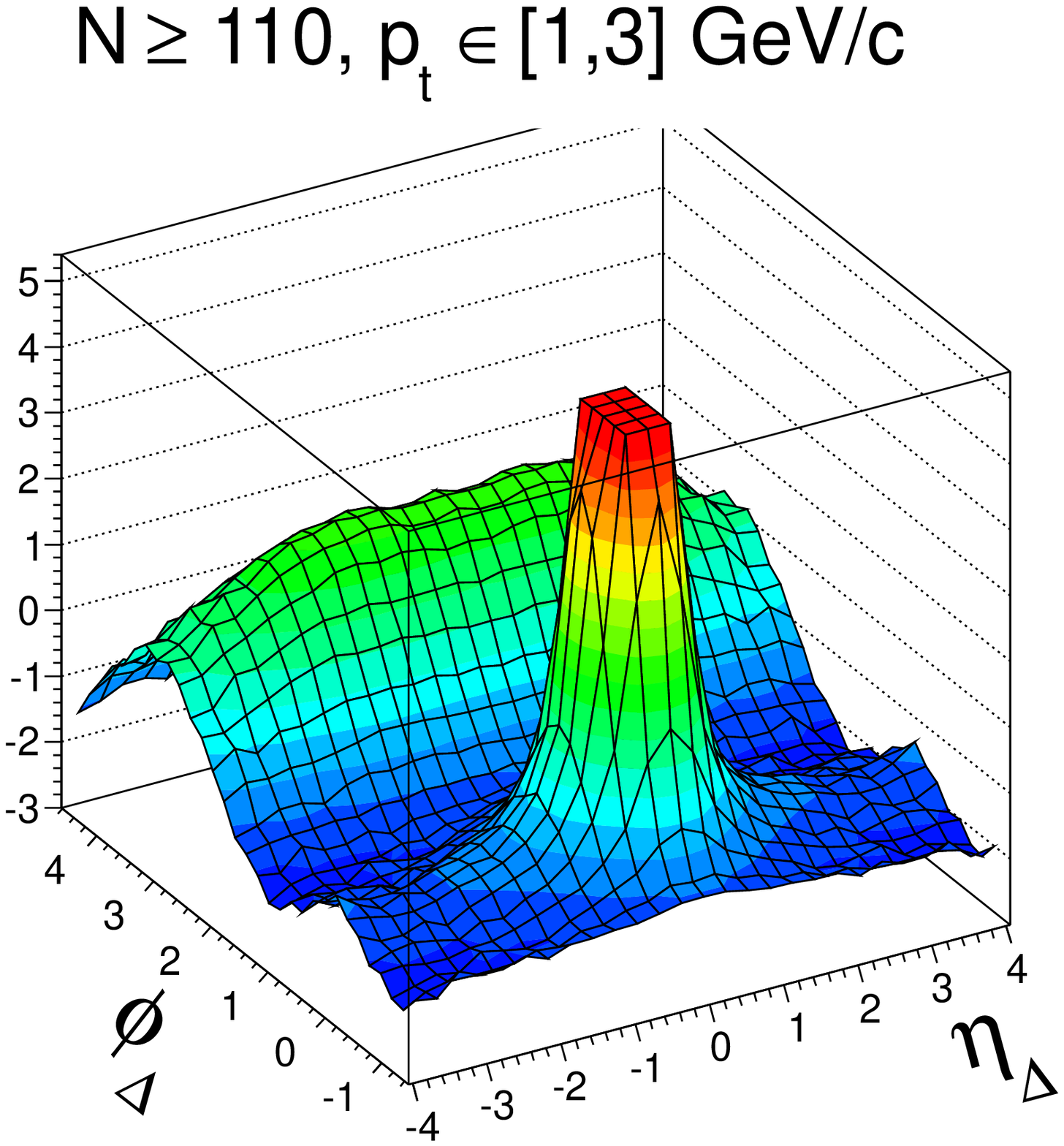}
\includegraphics[keepaspectratio,width=1.70in]{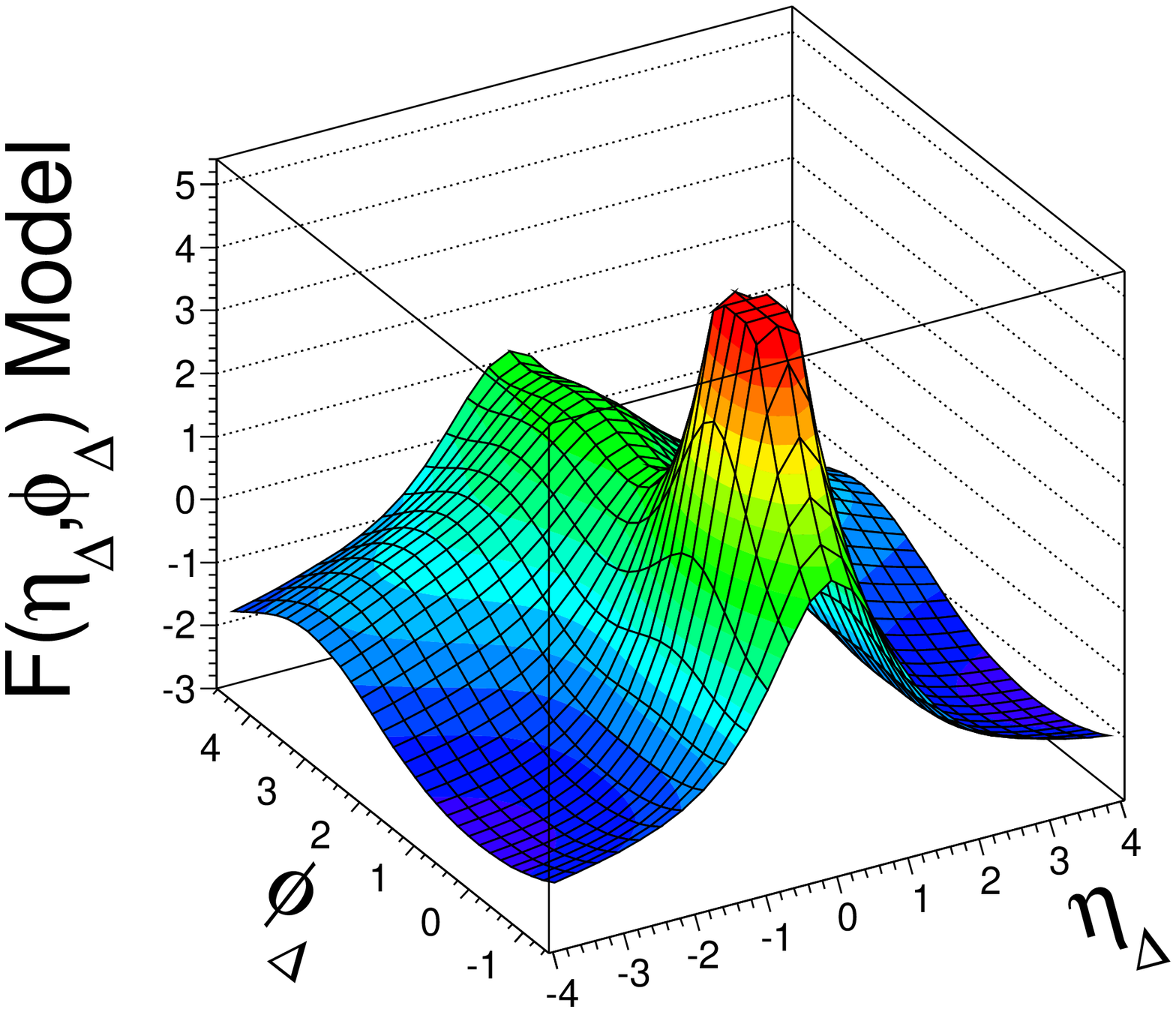}
\includegraphics[keepaspectratio,width=1.70in]{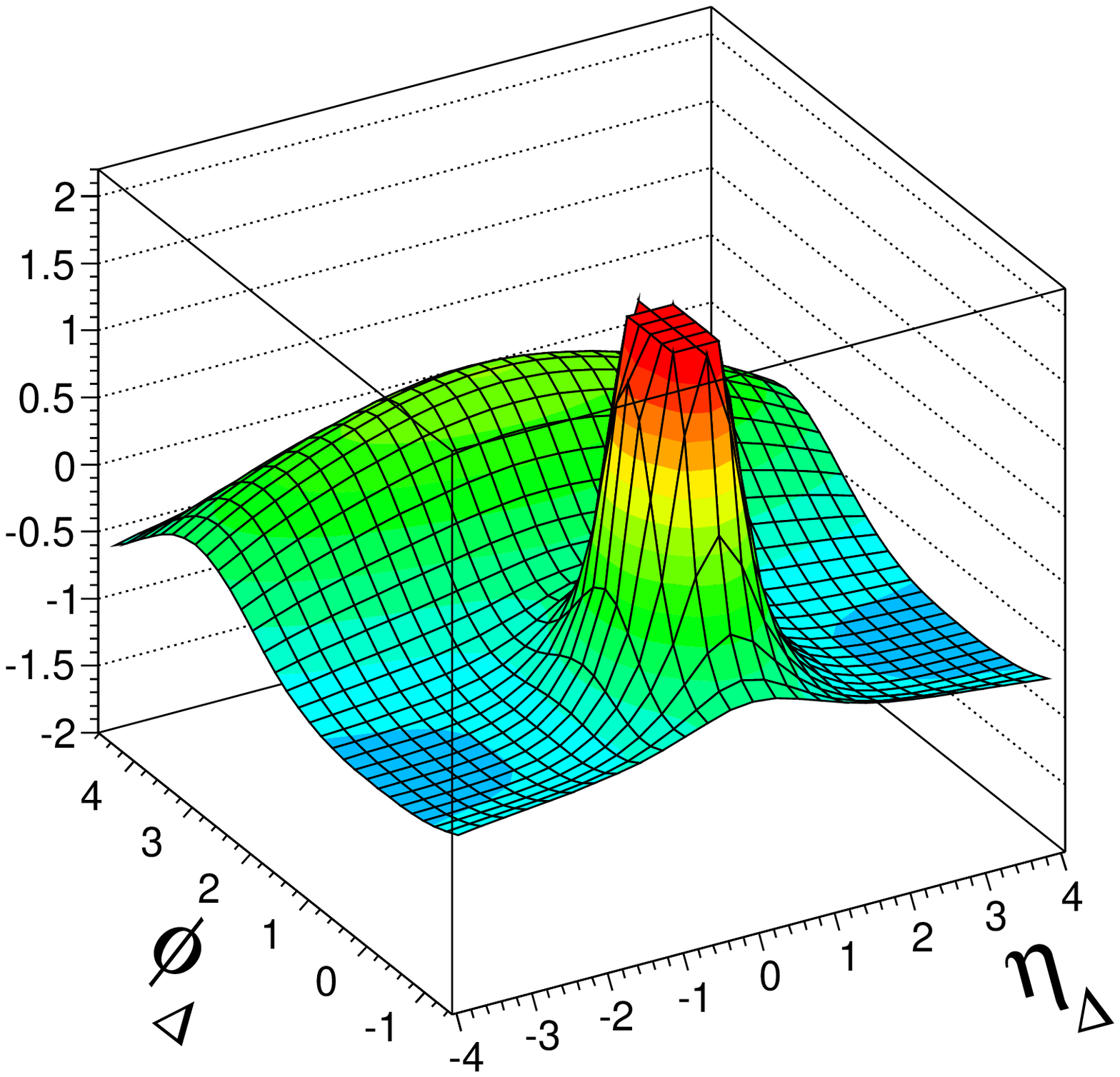}
\includegraphics[keepaspectratio,width=1.70in]{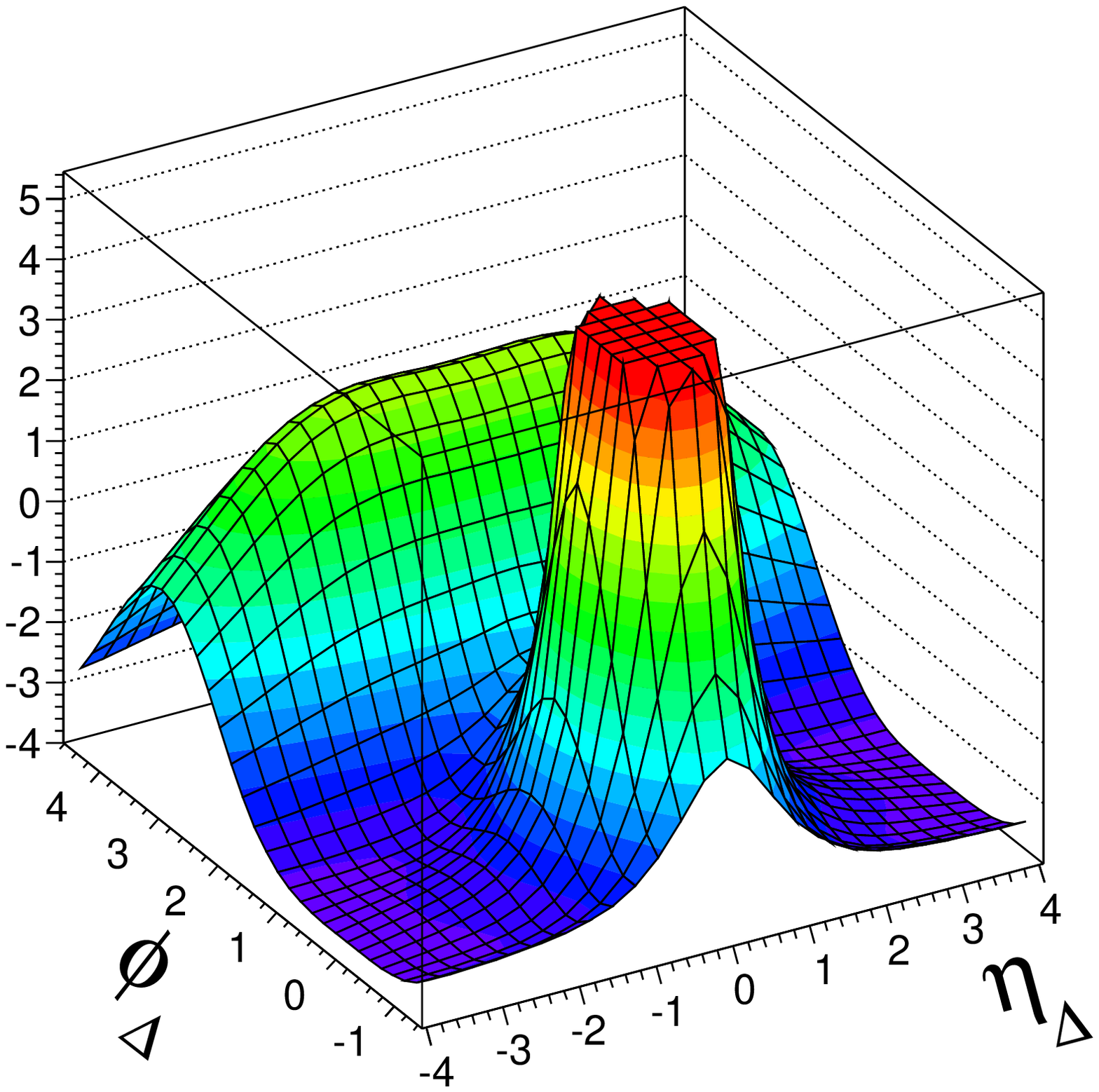}
\includegraphics[keepaspectratio,width=1.70in]{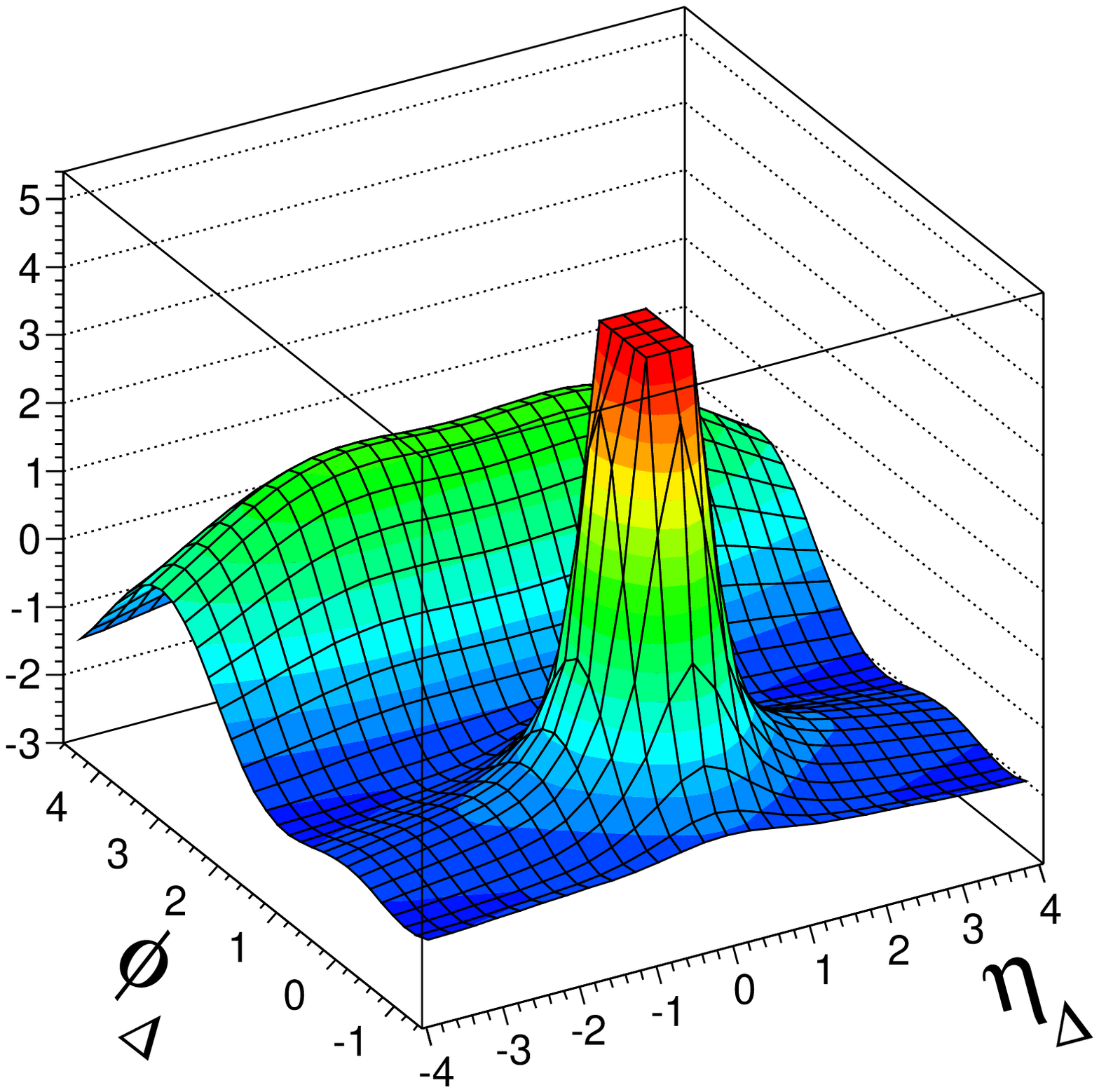}
\includegraphics[keepaspectratio,width=1.70in]{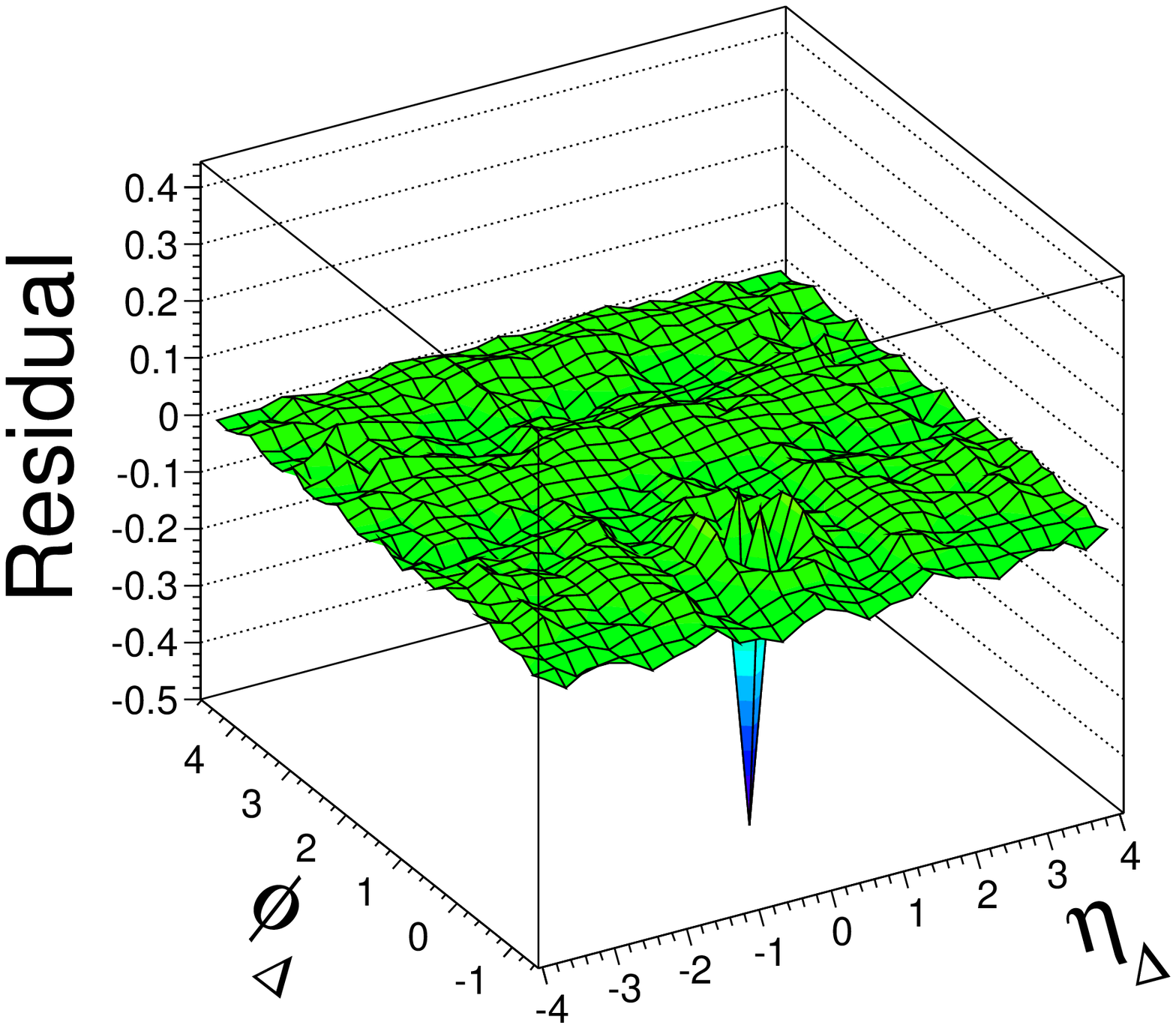}
\includegraphics[keepaspectratio,width=1.70in]{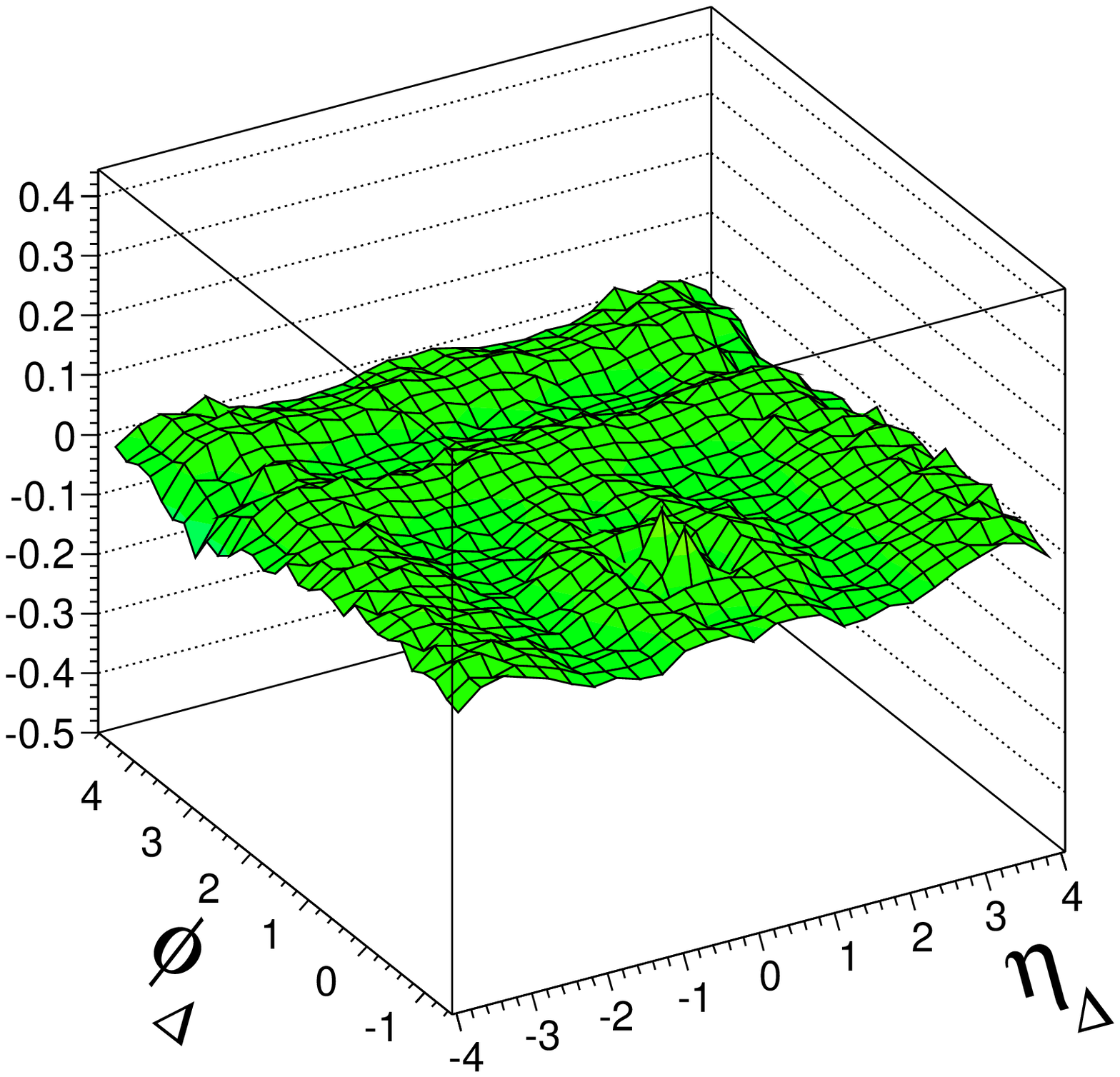}
\includegraphics[keepaspectratio,width=1.70in]{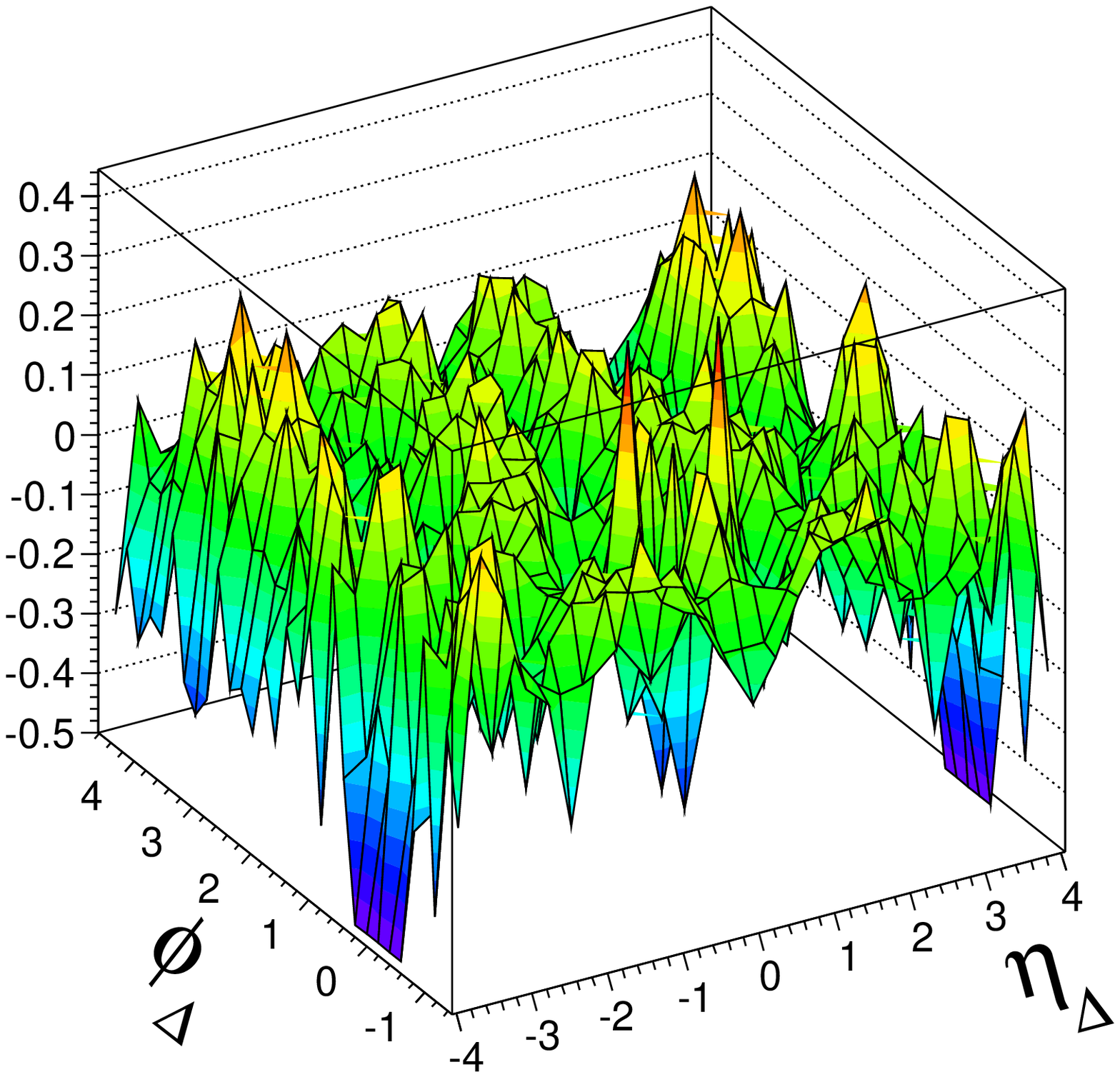}
\includegraphics[keepaspectratio,width=1.70in]{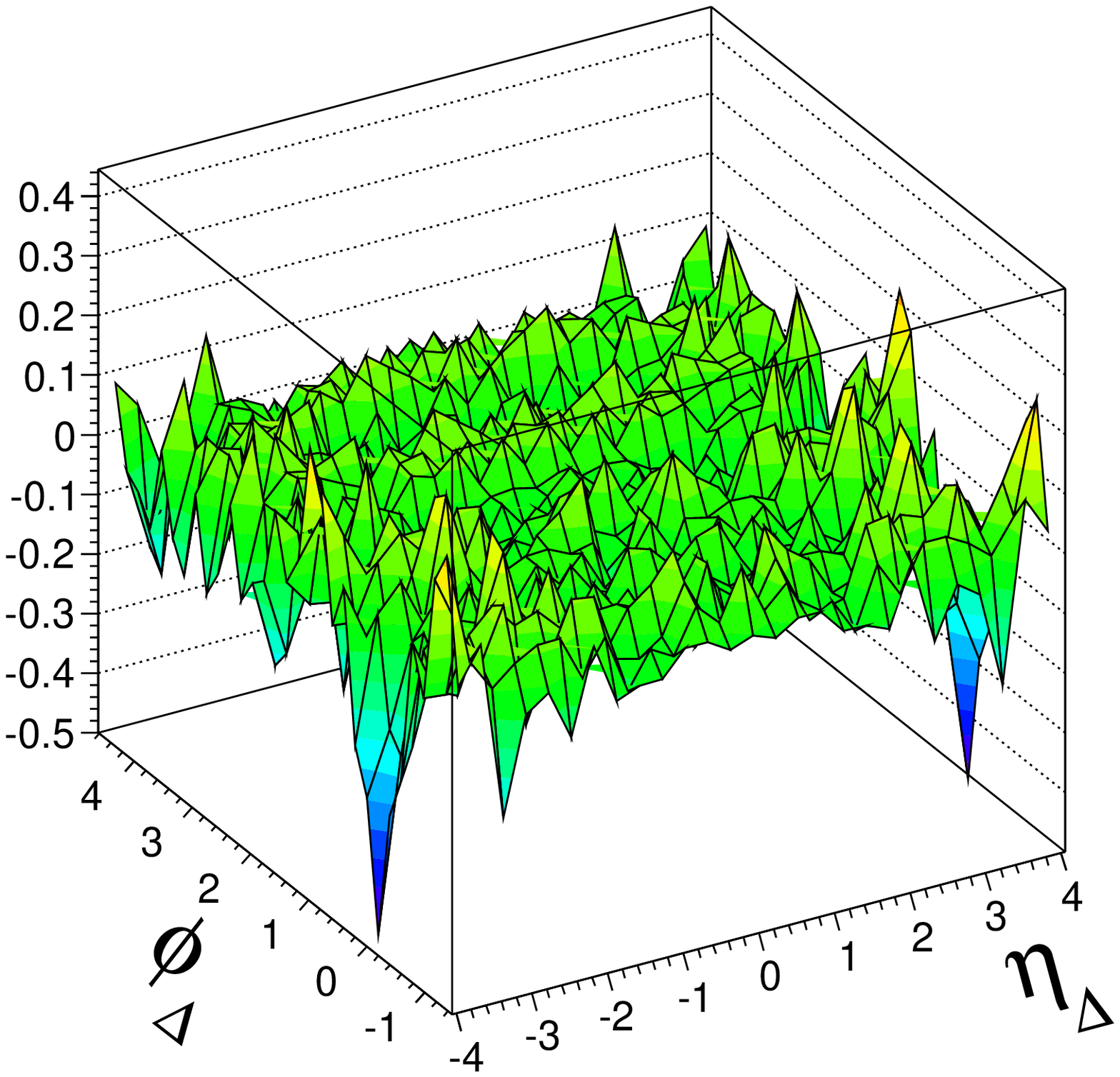}
\caption{\label{Fig1} 
(Color online) Perspective views of two-dimensional charge-independent correlations
for p-p minimum-bias and high event multiplicity ($N \geq 110$) collisions at $\sqrt{s}$ = 7~TeV from the CMS collaboration~\cite{CMS}. Upper, middle and lower rows of panels display data, model fits, and residuals (model - data), respectively. The columns of panels from left to right correspond to minimum-bias events with
charged particle selection $p_t > 0.1$~GeV/c and $1 < p_t < 3$~GeV/c, and
 high multiplicity events with $p_t > 0.1$~GeV/c and $1 < p_t < 3$~GeV/c, respectively.
The fitting models selected for display are described in the text. All residuals are shown using the same, expanded scale.}
\end{figure*}

The fitting results for the high event multiplicity data with
$1 < p_t < 3$~GeV/$c$ from the CMS experiment are discussed first.
Initially the data were fit using the $A_0$, $A_1$, $A_2$ and $A_4$
terms in Eq.~(\ref{Eq1}) where 1D Gaussian component $A_3$ was not required.
The same-side peak structure was accurately described by the combination of
approximate 2D
Gaussian and 2D exponential terms where the fitted exponents of the former term
equaled 2 (Gaussian) within 2\% (but note the caveat in Sec.~\ref{Sec:FitModel}). 
The away-side ridge was well described by
the periodic Gaussian ($A_4$ term) with moderate $\eta_\Delta$
modulation. Although the SS peak and AS ridge were well described the 
$\chi^2$ was rather large due to the inability of the model to account for the
$\eta_\Delta$-extended same-side ridge. 
Including
the soft gluon radiation correlations with adjustable, non-negative amplitude
failed to improve the fit quality in this instance. $\chi^2$/dof 
was improved significantly however (by 12 units) when either the quadrupole
or SS periodic Gaussian ridge ($A_5$ term) was included. Fits including the
quadrupole obtained the lowest $\chi^2$ (see Table~\ref{TableI}). The data,
the best fit (with quadrupole), and the residuals (model - data) are shown
in Fig.~\ref{Fig1} in the right-most column of panels.  The soft
gluon radiation correlations afforded very little improvement in fit quality for
the SS Gaussian ridge model, but significantly improved the $\chi^2$/dof from
4.18 to 4.04 when the quadrupole was included.

Similar results were found for the high multiplicity, $p_t > 0.1$~GeV/$c$
data except that the $\chi^2$ improvement obtained with either the additional
quadrupole or SS Gaussian was less dramatic (about 4 units for $\chi^2$/dof)
than was obtained for the higher $p_t$ data. The SS 2D
peak function ($A_1$) relaxed to a non-Gaussian geometry with both pseudorapidity
and azimuth exponents less than 1 (see Table~\ref{TableI}). Statistically
significant evidence for both the quadrupole and SS Gaussian ridge was found.
The best fit was obtained assuming the additional quadrupole component. The
soft gluon radiation correlations provided no significant improvement in the fit
quality. The data, best fit (quadrupole model), and residuals are shown
in Fig.~\ref{Fig1} in the third column of panels from the left.

The quadrupole and same-side Gaussian ridge can have very similar shapes and
amplitudes for the same-side azimuth range. The two are of course very
different on the away-side.
Evidently the $\chi^2$/dof reduction afforded by the quadrupole
for both high event multiplicity data sets is due, at least
in part, to improvements in the fit quality for the away-side data.
Visual evidence of an away-side quadrupole
contribution to the CMS high multiplicity data is not apparent, unlike the
angular correlation data for mid-central relativistic heavy ion
collisions~\cite{MikeQM08,DavidHQ}. Quantitative analysis of the type presented
here and in Refs.~\cite{Tomridge,Bozek} is required to differentiate
between these two descriptions of the SS and AS ridges.
Both the quadrupole and SS Gaussian amplitudes
increase for the higher $p_t$ selected particle cuts.
A modest, negative $\eta_\Delta^2$ modulation was obtained for both
the quadrupole and SS Gaussian which had only minor effects.

The magnitudes of the residuals for the high event multiplicity data are
less than 0.1 except near the origin and the
$\eta_\Delta$ acceptance edge and are about 1\% of the magnitudes of the principal correlation structures.
Although the residuals increase at large $|\eta_\Delta|$ the increasing statistical
errors  at large $|\eta_\Delta|$ prevent these bins from making significant $\chi^2$ contributions.
The band structures in the residuals on $\eta_\Delta$ and
$\phi_\Delta$ are not understood and could be of experimental and/or dynamical origin.
In any
case the residuals are small relative to the principal correlation structures including the quadrupole and SS
Gaussian ridge.

Both sets of minimum-bias correlation data were well described with just the
$A_0$, $A_1$, $A_2$, $A_3$ (only for the $p_t > 0.1$~GeV/$c$ data) and $A_4$
components. $\chi^2$ for the
$p_t > 0.1$~GeV/c data was dominated by the single (0,0) angular bin
which was subsequently omitted in the $\chi^2$ minimization procedure.
No evidence for a non-negative quadrupole or SS Gaussian ridge structure was
found. However, including the soft gluon radiation correlations improved
the $\chi^2$/dof by about 1 unit. The data, best fits (no quadrupole or SS
Gaussian ridge), and residuals are shown in the first two columns of panels
in Fig.~\ref{Fig1}.
The magnitude of the residuals is less than 0.03 except near the origin and the
$\eta_\Delta$ acceptance edge and is about 1\% of the magnitudes of the principal correlation structures.

%%%%%%%%%%%%%%%%%%%%%%%%%%%%%%%%%%
%%%%%%%%%%%%%%%%%%%%%%%%%%%%%%%%%%

%%%%%%%%%%%%%%%%%%%%%%%%%%%%%%%%%%
\begin{table*}[t]
\caption{Best fit model parameters with fitting statistical errors (in parentheses)
for p-p minimum-bias and high event multiplicity ($N \geq 110$) collisions at $\sqrt{s}$ = 7~TeV.}
 \label{TableI}
\begin{tabular}{ccccccc}
\hline \hline
Parameter & MB $p_t>0.1$~GeV/$c$ & MB $p_t \in [1,3]$~GeV/$c$ & \multicolumn{2}{c}{$N\geq 110,p_t>0.1$~GeV/$c$} &
\multicolumn{2}{c}{$N\geq 110,p_t \in [1,3]$~GeV/$c$} \\
          & Jet+ASG     &   Jet+ASG  & Quad  & SSG  & Quad  &  SSG \\
\hline
$A_1$     & 1.915(19)  & 4.40(3)  & 11.63(15)  & 11.90(17)  & 16.52(15)  & 16.17(14)  \\
$\sigma_{\eta_\Delta}$ & 0.501(3) & 0.313(1) & 0.426(4) & 0.422(4) & 0.285(1) & 0.285(1) \\
$\alpha$ & 1.021(5) & 1.032(4) & 0.938(8) & 0.916(7) & 0.979(5) & 0.990(5) \\
$\sigma_{\phi_\Delta}$ & 0.737(7) & 0.370(1) & 0.422(7) & 0.475(6) & 0.310(1) & 0.318(2) \\
$\beta$ & 0.883(9) & 1.035(4) & 0.740(6) & 0.794(7) & 1.003(6) & 1.012(7) \\
\hline
$A_2$  & 8.07(3) & 6.88(3) & 35.0(2) & 34.8(2) & 14.3(2) & 14.68(15) \\
$w_\eta$ & 0.753(2) & 0.462(1) & 0.446(2) & 0.446(3) & 0.418(3) & 0.421(3) \\
$w_\phi$ & 0.807(5) & 0.429(1) & 0.458(3) & 0.435(2) & 0.377(2) & 0.363(2) \\
\hline
$A_3$  &  1.532(24) & $-$  & $-$ & $-$ & $-$ & $-$ \\
$\sigma_0$ & 0.826(3) & $-$  & $-$ & $-$ & $-$ & $-$ \\
$\gamma$ & 0.931(3) & $-$  & $-$ & $-$ & $-$ & $-$ \\
\hline
$A_4$ & 2.125(12) & 1.446(1) & 5.62(2) & 6.23(3) & 3.08(2) & 3.80(2) \\
$\sigma_{AS}$ & 1.196(2) & 0.956(1) & 0.947(8) & 0.862(4) & 1.156(31) & 0.841(5) \\
$\epsilon$ & -0.021(1) & -0.0326(1) & -0.0231(3) & -0.0242(3) & -0.020(1) & -0.0208(3) \\
$\zeta$  & -0.055(3) & 0.0080(6) & -0.040(1) & -0.035(1) & -0.046(2) & -0.042(1) \\
\hline
$A_Q$ & $-$  & $-$ & 0.486(23)\footnote{Additional 15\% systematic uncertainty should be included due to quoted systematic error in the ridge signal~\cite{CMS}.} & $-$ & 0.657(24)$^a$ & $-$ \\
$A_5$ & $-$ & $-$ & $-$ & 0.556(40)$^a$ & $-$ & 0.846(27)$^a$ \\
$\sigma_{SS}$ & $-$ & $-$ & $-$ & 0.550(28) & $-$ & 0.583(15) \\
$\delta$ & $-$  & $-$ & -0.056(2) & -0.017(6) & -0.032(1) & -0.021(3) \\
\hline
$A_0$ & -1.344(5) & -0.124(1) & -1.038(18) & -0.571(17) & -0.505(19) & 0.164(11) \\
$\chi^2$/dof & 14.3 & 16.8 & 9.78 & 10.9 & 4.18 & 4.57 \\
Vol(2DG) & 4.64 & 3.13 & 15.5 & 17.2 & 9.23 & 9.20 \\
$A_4\sigma_{AS}$ & 2.54 & 1.38 & 5.32 & 5.37 & 3.56 & 3.20 \\
$v_2$  & $-$  & $-$ & 0.046 & $-$ & 0.053 & $-$ \\
\hline \hline
\end{tabular}
\end{table*}
%%%%%%%%%%%%%%%%%%%%%%%%%%%%%%%

For each case the same-side peak structure
was well described with a combination of approximate 2D Gaussian
and 2D exponential functions. The exponent fit parameters $\alpha$ and $\beta$
were generally close to unity corresponding to a 2D Gaussian-like peak shape.
For each data set the five parameters of the 2D Gaussian-like component
were not significantly affected by including or not including the quadrupole
and SS Gaussian ridge.
%The amplitude of the 2D Gaussian-like portion
%of the same-side 2D correlation peak,
%in general
%follows trends expected for fragments from minimum-bias jets (i.e. jets with no
%minimum $p_t$ or energy requirement)~\cite{Tomfrag}.
Trainor and Kettler~\cite{Tomridge} showed that
the minimum-bias jet correlation structure in the CMS data is
similar to that at RHIC energies when scaled by $\log(\sqrt{s})$.
The total number of jet-like correlated pairs per final-state
particle, estimated by the volume of the same-side 2D Gaussian-like peak (see 
Table~\ref{TableI}, but note the caveat in Sec.~\ref{Sec:FitModel}), increases for high multiplicity events because those
events have increased probability of minimum-bias jet production within the
acceptance. However, this volume decreases for the higher $p_t$
($1 < p_t < 3$~GeV/c) particle selection cut.  Studies of 2D
charged particle $(p_{t1},p_{t2})$ correlations from $\sqrt{s}$ = 0.2~TeV
minimum-bias NSD p-p collisions at RHIC~\cite{Jeffpp} suggest that the
$p_t \in [1,3]$~GeV/$c$ cut-range
may exclude a large fraction of the correlated pairs
produced by minimum-bias jets. This may account for some of the
reduction. The widths of the same-side jet-like peak decrease
with higher $p_t$ selection as expected for jet fragmentation with uniform transverse momentum
distributions relative to the jet thrust axis~\cite{Jetfrag}.
Further analysis of the minimum-bias jet-like correlations
in p-p collisions at the LHC will
likely require different $p_t$ selection cuts which are optimized
for that purpose.

In general the away-side
azimuth ridge structure was better described with an $\eta_\Delta$-dependent, periodic
Gaussian than with an $\eta_\Delta$-dependent dipole.
A periodic series of 1D Gaussians, as in Eq.~(\ref{Eq1}), approaches an azimuth
dipole plus offset for sufficiently large width $\sigma_{AS}$ ({\em e.g.}
higher-order multipole contributions are $<$ 1\% for $\sigma_{AS} > 1.75$~\cite{footperiod}).
The fitted widths here are less than 1.2 indicating that the dipole limit is not
reached.
The amplitude $A_4$ approximately follows 
the volume trends of the SS 2D peak as expected if this correlation
structure is dominated by inter-jet charged particle fragment pairs for dijets
within the tracking acceptance. The majority of the $\eta_\Delta$ dependence
(fall-off) of the away-side ridge (see Fig.~\ref{Fig1}) is described by the
negative $\epsilon \eta_\Delta^2$ modulation. Weaker dependences
were approximated by the $\zeta \cos(2\pi \eta_\Delta / \Delta \eta)$
modulation. Model descriptions of the AS $\eta_\Delta$ dependence did not
significantly affect the fitted parameters of the SS 2D peak or the SS ridge
structure.
The away-side Gaussian decreased in amplitude and increased in width when the
quadrupole component was included in the fitting.

%%%%%%%%%%%%%%%%%%%%%%%%%%%%%%%%%%%%%%%%%%%%%%%%%%%%%%%%%%%%%%%%%%%%%%%%%%%
\section{Discussion}
\label{Sec:IV}
%%%%%%%%%%%%%%%%%%%%%%%%%%%%%%%%%%%%%%%%%%%%%%%%%%%%%%%%%%%%%%%%%%%%%%%%%%%

The CMS collaboration noted the similarity between the same-side $\eta_\Delta$
elongated correlations and that reported for Au-Au collisions
at RHIC by the STAR~\cite{Ayaminijet,MikeQM08,JoernRidge} and
PHOBOS~\cite{PHOBOSRidge} collaborations
for total energy per nucleon-nucleon (N-N) pair
$\sqrt{s_{NN}}$ = 0.2~TeV.
What was observed in minimum-bias
collisions at RHIC using all charged particles with $|\eta| \leq 1.0$ and
$p_t > 0.15$~GeV/c was a 2D approximately Gaussian peak
centered at $(\eta_\Delta,\phi_\Delta) = (0,0)$
whose amplitude and
$\eta$ width increase substantially and monotonically with decreasing impact
parameter between the colliding ions (increasing centrality).
In addition analysis of these data provided evidence for an $\eta_\Delta$-independent
quadrupole which smoothly varies with collision centrality and is
approximately proportional to the square of the spatial eccentricity of the
overlapping, colliding ions.
In similar correlation studies
using higher momentum ``trigger'' particles with lower momentum
associated particles the same-side peak evolved to a non-Gaussian
structure~\cite{JoernRidge}. It is plausible that jet fragmentation could be
modified in the dense environment produced in relativistic
heavy ion collisions~\cite{Borghini}.

Field~\cite{Field} pointed out that initial and final
state radiative processes ($2 \rightarrow 3$, $2 \rightarrow 4$)
accompanying transverse partonic scattering preferentially produce hadronic
fragments in the hard-scattering plane defined by the beam and jet axes.
Single gluon radiation in $2 \rightarrow 2$ hard processes was calculated
in the soft gluon limit in~\cite{Ellis,Jaques}. Using the $gg \rightarrow ggg$
distribution from \cite{Ellis} additional angular correlations for the
p-p collision system were calculated and included in the present fitting model. Although
the additional correlations made statistically significant improvements in the
fit quality for the minimum-bias collision data, they could not explain the same-side
$\eta_\Delta$-extended ridge in the high event multiplicity data. 

Sj\"{o}strand~\cite{Sjostrand} and Trainor~\cite{Tomfrag}
suggested that color-flux tube connections between transverse (scattered) and
longitudinal (beam direction) partons could, in analogy to the LUND color-string fragmentation
model~\cite{LUND} for soft particle production, produce an $\eta_\Delta$
extended, exponentially decreasing enhancement in the jet-like correlations.
The present analysis does not find evidence of increasing $\eta$ width
(i.e. increased value of $\sigma_{\eta_\Delta}$) or evolution toward an exponential
shape [exponent parameter $\alpha$ in Eq.~(\ref{Eq1})] in conjunction with the
appearance of the same-side ridge, in agreement with previous findings
by Trainor and Kettler~\cite{Tomridge}.
Nevertheless, this suggested jet fragmentation mechanism 
should be further considered with respect to
the $\eta_\Delta$-elongated correlations observed in A-A collisions
at RHIC and the LHC~\cite{Ant}.

Two classes of theoretical models purport to explain the same-side
$\eta_\Delta$-elongated ridge correlations observed in the RHIC heavy
ion data. The first assumes event-by-event energy and/or
momentum density fluctuations in the initial collision stage immediately after
impact which propagate outward via pressure driven radial
expansion.
The initial stage fluctuations are described as 
color-glass condensate (CGC)~\cite{CGC} flux tubes (glasma)~\cite{Dumitru},
localized ``hot spots'' in a quark-gluon plasma~\cite{Werner,Gavin,Andrade}, or beam-jet
remnants from hard scattering processes~\cite{Shuryak,Voloshin}. The second
class of models assumes hard, transverse parton $-$ soft, longitudinal parton
interactions via recombination~\cite{Hwa} or strong rescattering~\cite{Wong}
to induce the $\eta_\Delta$-elongated correlations.

So far, most proponents have
only attempted to address a subset of the experimental features associated
with the same-side
angular correlations, {\em e.g.} either the width or amplitude dependence
on collision centrality.
Accompanying correlation
structures including the away-side ridge, charge-dependence (like - unlike
charged pair difference), and the 2D two-particle $(p_{t1},p_{t2})$
correlation~\cite{RayCERP} have not yet been addressed by these theoretical
models.
A recent analysis~\cite{RayCGC} of $(p_{t1},p_{t2})$ correlation predictions
for CGC flux tube induced correlations~\cite{Lappi}
shows that such initial stage models,
even including strong radial flow, are irrelevant for describing the same-side
$\eta_\Delta$-extended correlations.
Nevertheless, these models
motivated the periodic, same-side Gaussian ridge component adopted in the present
phenomenological study. Fits to the minimum-bias data provide no significant
evidence for this component while fits to the high event multiplicity,
$N \geq 110$ data do.

The quadrupole angular correlation continues to be extensively studied in
relativistic heavy ion collisions~\cite{Tomv2bash} and was recently
reported for Pb-Pb collisions at the LHC by the ALICE collaboration~\cite{ALICEflow}.
The most compelling explanation for the quadrupole correlation
is that it is a consequence of
anisotropic particle emission of the form $[1 + 2 v_2 \cos(2(\phi -
\psi))]$, where $\psi$ is a random, event-wise angle for maximum azimuth
particle density.
Averaging over an event ensemble
produces two-particle correlation component, $2 v_2^2 \cos(2\phi_\Delta)$.
The anisotropic emission presumably reflects the initial geometry (i.e.
eccentricity $\epsilon$) of the transverse overlap region between the 
colliding constituents of the two nuclei. Measurements show that observed
$v_2$ is approximately proportional to initial eccentricity estimated using
a Glauber model~\cite{DavidHQ}. Naive
superposition of independent N-N collisions, even if the spatial distribution of
N-N collision positions is non-isotropic, produces on average an isotropic
final-state particle distribution on azimuth resulting in $v_2 = 0$.
What causes the azimuth modulation ($v_2$)
is a mystery although theories are abundant (e.g. \cite{hydro} but note
discussion in~\cite{Tomv2bash}).

Trainor and Kettler~\cite{Tomridge} and Bo\.zek~\cite{Bozek} both showed that the p-p
high multiplicity 7~TeV CMS correlation data could be well described with a quadrupole component.
Trainor and Kettler described
the 7 TeV CMS data with a similar, 2D fitting model as in Eq.~(\ref{Eq1})
but without the 2D exponential, same-side Gaussian ridge, and $\eta_\Delta$
modulations assumed here. 
They showed that the quadrupole amplitude inferred for the
7~TeV high multiplicity p-p correlation data is consistent with the simple,
linear trends on collision energy [proportional to $\log(\sqrt{s_{NN}})$], eccentricity and
event multiplicity reported for Au-Au minimum-bias collisions at RHIC~\cite{DavidHQ}.
Bo\.zek~\cite{Bozek} fitted the 1D azimuth
projections for $|\eta_\Delta|$ from 2.0 to 4.8 for sixteen combinations of
event multiplicity and $p_t$ interval~\cite{CMS} using $\cos(\phi_\Delta)$ and
$\cos(2\phi_\Delta)$ terms only. Good fits were obtained although contributions
of the same-side jet peak for $|\eta_\Delta| > 2$ occur 
and the azimuth projection averages over the strong $\eta_\Delta$ dependence
of the away-side ridge beginning at $|\eta_\Delta| > 2$.
The quadrupole amplitude $A_Q$ in Eq.~(\ref{Eq1}) for the CMS correlation
quantity $R$ (see Eq.~(1) in \cite{CMS}) is related to azimuth anisotropy parameter
$v_2$ defined above by,
\bea
A_Q & = & 2 \left[ \langle N^{offline}_{trk} \rangle -1 \right] v^2_2.
\label{Eq3}
\eea
Inferred values of $v_2$ from the present analysis (see Table~\ref{TableI}) and
Eq.~(\ref{Eq3}) are $v_2 = 0.046$ and 0.053 for the $N \geq 110$, $p_t > 0.1$~GeV/c
and $1 < p_t < 3$~GeV/c data, respectively, in fair agreement with
$\sim 0.04$ obtained in both Trainor and Kettler~\cite{Tomridge} and Bo\.zek~\cite{Bozek}. The best fits obtained in this analysis were those
with the quadrupole component as discussed in Sec.~\ref{Sec:III}.

The CMS experiment also reported strong variation in the large
$|\eta_\Delta|$ ($2.0 < |\eta_\Delta| < 4.8$) same-side ridge amplitude for high
multiplicity events ($N>90$) in four $p_t$ ranges from 0.1 to 4~GeV/c (see Fig.~9 in \cite{CMS}).
The larger values occurred in the $p_t \in [1,2]$ and $[2,3]$~GeV/$c$ bins.
A similar increasing then decreasing trend with $p_t$ was reported by the STAR experiment~\cite{DavidHQ}
for the Au-Au quadrupole correlation
in more central Au-Au collisions.
CMS also found that the same-side ridge correlations
for like-sign and unlike-sign charged particle pairs were the same within
uncertainties as shown in Fig.~10 of \cite{CMS}. Quadrupole correlations in Au-Au collisions at RHIC are also
charge-independent within errors~\cite{Ayaminijet,MikeQM08,DavidHQ}.
Non-zero charge-dependent angular
correlations in Au-Au collisions at RHIC extend out to only 2 units on
relative $\eta$~\cite{AyaCD}. The CMS results in
Figs.~9 and 10 of Ref.~\cite{CMS} are qualitatively consistent with quadrupole
phenomenology for Au-Au collisions at RHIC energies.

The conventional assumption in the heavy ion physics literature is that strong
interactions among the (partonic) constituents lead to rapid thermal
equilibration within about 1~fm/$c$ after impact~\cite{Mueller}.
This results in pressure gradients which drive
collective expansion as described by particle transport
models~\cite{PCM} or
hydrodynamic evolution~\cite{hydro}. Hadronization models are invoked in
the final stages of expansion which enable predictions to be compared
with experiment.
%Those models approximately describe the observed single
%particle $p_t$ spectrum and $v_2$
%values within limited kinematic ranges
%(low $p_t$ and $|\eta|$)~\cite{flowsummary}.
Event-wise fluctuations in energy and momentum density in the
initial stage of the collision are expected to propagate to the final state
producing correlations.
Critical to this interpretation is the requirement of very rapid thermal
equilibration; delays beyond a few fm/c result in insufficient hydrodynamic pressure to
account for observed $v_2$ values.
It has been shown~\cite{Kovchegov} that rapid thermal equilibration cannot
occur via pQCD processes or by non-perturbative instanton mechanisms in the brief
amount of time available (of order 10~fm/$c$) during the heavy ion collision
process. That thermal equilibration might occur during a p-p collision is
even more problematic due to the reduced spatial and temporal scales involved.
Kovchegov~\cite{Kovchegov} concluded that if
rapid thermal equilibration does occur some as yet unidentified,
non-perturbative
mechanism would be required.
%Predictions
%of minijet thermalization in heavy ion collisions~\cite{Mueller,XinNian,?},
%an unavoidable consequence of the above strong interactions necessary
%to produce rapid equilibration, does not appear to occur, even in central Au-Au
%collisions at full RHIC energy~\cite{minijets-hydro}.
Therefore, if the
p-p CMS correlation data do in fact display a quadrupole correlation as
concluded in~\cite{Tomridge,Bozek} and indicated in the present results via
$\chi^2$ selection, a production mechanism for the quadrupole correlation other than rapid thermal
equilibration is implied.

Three alternate mechanisms to the hydrodynamic scenario for generating the azimuth modulation in heavy
ion collisions have been proposed. Boreskov {\em et al.}~\cite{Boreskov}
assumed a non-local N-N inclusive collision vertex model with final-state
emission oriented relative to the momentum transfer. Convolution of this
vertex function with the transverse nucleus-nucleus overlap distribution,
which includes important anisotropic density gradients, produced anisotropic final-state
particle distributions with $v_2$ values of reasonable magnitude. Kopeliovich {\em et al.}~\cite{Kopeliovich}
used a color-dipole model and predicted $v_2$ for p-p collisions at 200 GeV of
order a few percent, which is a reasonable magnitude given the present results and those
in~\cite{Tomridge,Bozek} and the empirical energy dependence of the quadrupole
correlation reported by Kettler~\cite{DavidHQ}.
Trainor~\cite{Tomv2bash} suggested that interacting color
currents in the nucleus-nucleus overlap region could emit gluon multipole
radiation in analogy to classical electrodynamics.

%%%%%%%%%%%%%%%%%%%%%%%%%%%%%%%%%%%%%%%%%%%%%%%%%%%%%%%%%%%%%%%%%%%%%%%%%%%
\section{Summary and Conclusions}
\label{Sec:V}
%%%%%%%%%%%%%%%%%%%%%%%%%%%%%%%%%%%%%%%%%%%%%%%%%%%%%%%%%%%%%%%%%%%%%%%%%%%

Theoretically motivated phenomenological fitting models were applied to the
two-particle 2D angular correlation data for p-p collisions at $\sqrt{s} = 7$~TeV
reported by the CMS Collaboration.  The dominant features of the
data were well described by jet-like correlation structures
projected onto relative $\eta,\phi$ subspace, consisting of a 2D Gaussian-like peak
at the origin with accompanying away-side ridge at $\phi_\Delta = \pi$ representing pair
correlations between dijets within the acceptance. For the high multiplicity
events with $N \geq 110$ a new, extended correlation ridge along $\eta_\Delta$ 
was observed at
small relative azimuth.
The same-side jet-like component was allowed to have (1) an extended width along
$\eta_\Delta$ which is required to describe similar angular correlation data 
from Au-Au collisions at RHIC, (2) non-Gaussian distortions which are 
suggested by possible higher-order fragmentation processes~\cite{Sjostrand,Tomfrag},
and (3) additional soft radiated gluon - jet particle correlations~\cite{Ellis,Jaques}.
None of these extensions enabled the jet related components to describe the
same-side correlation ridge.
%A fitting model with extended width along $\eta_\Delta$ of the jet-like
%correlation peak, which was motivated by
%observations of $\eta_\Delta$-extended same-side angular correlations in Au-Au
%collision data from RHIC, non-Gaussian distortions of the jet-like peak
%suggested by higher-order processes~\cite{Sjostrand,Tomfrag}, and additional
%soft radiated gluon - jet particle correlations~\cite{Ellis,Jaques},
%were
%unable to describe this new feature.

However, the same-side ridge was well described 
when either an additional same-side Gaussian ridge on
azimuth or an azimuth quadrupole $[\cos(2\phi_\Delta)]$ was included in the
fitting. The former structure was predicted for dense, strongly interacting
systems where initial energy/momentum density fluctuations are propagated
to the final hadronic state via pressure driven radial flow. The latter
(quadrupole) has been observed in angular correlations for relativistic heavy ion
collisions.

Based on the present analysis neither the quadrupole or same-side Gaussian
ridge phenomenological
descriptions can be ruled out, although the quadrupole model results in smaller
$\chi^2$/dof and residuals for both the lower $p_t$ ($p_t > 0.1$~GeV/$c$) and higher
$p_t$ ($1 < p_t < 3$~GeV/$c$) high event multiplicity data.
No evidence for either the quadrupole or same-side Gaussian ridge was found
in this analysis for the 7~TeV minimum-bias p-p correlation data.
Alternative mechanisms for producing the quadrupole correlation
have been proposed~\cite{Tomv2bash,Boreskov,Kopeliovich} which do not
rely on rapid thermal equilibration and pressure driven collective flow.
It would be interesting to apply those models to the CMS p-p correlation data.

The present analysis shows
that the CMS ridge could be a manifestation of the quadrupole correlation, well-known from
relativistic heavy ion collisions, and possibly appearing in p-p collisions for
the first time in these data from CMS. The phenomenological results presented here and in 
\cite{Tomridge,Bozek} together with the problematic assumption of rapid
equilibration in p-p collision systems, warrant further study of the underlying
mechanism(s) producing the quadrupole correlation in high energy collisions.

\vspace{0.1in}

The author expresses sincere thanks to Professors Guido Tonelli,
Roberto Tenchini, Gunther Roland and Dr. Wei Li of the CMS Collaboration
for providing the data used in this analysis and to Dr. Sasha Pranko for
helpful advice regarding soft gluon radiation in perturbative QCD.
I also thank Professors
Thomas Trainor and Duncan Prindle for many useful conversations and
suggestions regarding this and related two-particle correlation analysis
and 2D fitting over the years.  This work was supported in part by
The United States Department of Energy under grant No. DE-FG02-94ER40845.

%%%%%%%%%%%%%%%%%%%%%%%%%%%%%%%%%%%%%%%%%%%%%%%%%%%%%%%%%%%%%%%%%%%%%%%%%%%


\begin{thebibliography}{99}
\bibitem{footperiod}
A periodic series of 1D Gaussians on azimuth at odd multiples of $\pi$
approaches a constant plus dipole
according to,
$A\sum_{k=\pm{\rm odd-integ}} e^{-(\phi_\Delta - k\pi)^2/2\sigma^2} 
= A\sigma/\sqrt{2\pi} [ 1 + 2\sum_{m=1}^\infty (-1)^m
e^{-m^2\sigma^2/2} \cos(m\phi_\Delta)]$
which limits to
$A\sigma/\sqrt{2\pi} [1 - 2 e^{-\sigma^2/2} \cos\phi_\Delta \cdots]$
for $\sigma \gg 1$.
Similarly, a periodic series of 1D Gaussians at integer multiples of $\pi$, represented in
this notation as $2A\sigma/\sqrt{2\pi} [ 1 + 2\sum_{m=2,4\cdots}^\infty e^{-m^2\sigma^2/2} \cos(m\phi_\Delta)]$, limits to $2A\sigma/\sqrt{2\pi} [1 +2 e^{-2\sigma^2} \cos(2\phi_\Delta) \cdots]$ for $\sigma \gg 1$.
\bibitem{footambig}
In general, fitting functions with non-orthogonal components
are susceptible
to discrete and continuous fitting ambiguities. The only such ambiguity encountered
here was a discrete ambiguity between the same-side 2D Gaussian peak and
2D exponential when the exponent (1/2) of the latter was allowed to vary.
Restricting it to 1/2 eliminated the ambiguity.
\end{thebibliography}
\end{document}